\documentclass[prb,twocolumn,superscriptaddress,noshowpacs,notitlepage,longbibliography]{revtex4-1}%
\usepackage{graphicx,bm,times}
\usepackage{amsmath}
\usepackage{amsfonts}
\usepackage{amssymb}
\usepackage{color}
\usepackage{hyperref}
\hypersetup{colorlinks = true,	allcolors = {blue}}
\begin{document}
\title{Holstein polarons, Rashba-like spin splitting and Ising superconductivity in electron-doped MoSe$_2$}

\author{Sung~Won~Jung}
\email{sungwon.jung@gnu.ac.kr}
\affiliation{Diamond Light Source, Harwell Science and Innovation Campus, Didcot, OX11 0DE, United Kingdom}
\affiliation{Department of Physics and Research Institute of Molecular Alchemy, Gyeongsang National University, Jinju 52828, Republic of Korea}

\author{Saumya Mukherjee}
\affiliation{Diamond Light Source, Harwell Science and Innovation Campus, Didcot, OX11 0DE, United Kingdom}
\affiliation{Van der Waals–Zeeman Institute, Institute of Physics, University of Amsterdam, 1098 XH, Amsterdam, Netherlands}

\author{Matthew~D.~Watson}
\affiliation{Diamond Light Source, Harwell Science and Innovation Campus, Didcot, OX11 0DE, United Kingdom}

\author{Daniil~V.~Evtushinsky}
\affiliation{\'{E}cole Polytechnique F\'{e}d\'{e}rale de Lausanne, CH-1015 Lausanne, Switzerland}
	
\author{Cephise Cacho}
\affiliation{Diamond Light Source, Harwell Science and Innovation Campus, Didcot, OX11 0DE, United Kingdom}

\author{Edoardo Martino}
\affiliation{\'{E}cole Polytechnique F\'{e}d\'{e}rale de Lausanne, CH-1015 Lausanne, Switzerland}

\author{Helmut Berger}
\affiliation{\'{E}cole Polytechnique F\'{e}d\'{e}rale de Lausanne, CH-1015 Lausanne, Switzerland}

\author{Timur~K.~Kim}
\email{timur.kim@diamond.ac.uk}
\affiliation{Diamond Light Source, Harwell Science and Innovation Campus, Didcot, OX11 0DE, United Kingdom}

\date{\today} 

\begin{abstract}
Interaction between electrons and phonons in solids is a key effect defining physical properties of materials such as electrical and thermal conductivity. 
In transitional metal dichalcogenides (TMDCs) the electron-phonon coupling results in the creation of polarons, quasiparticles that manifest themselves as discrete features in the electronic spectral function. 
In this study, we report the formation of polarons at the alkali-dosed MoSe$_2$ surface, where Rashba-like spin splitting of the conduction band states is caused by an inversion-symmetry breaking electric field.
In addition, we observe the crossover from phonon-like to plasmon-like polaronic spectral features at MoSe$_2$ surface with increasing doping.
Our findings support the concept of electron-phonon coupling mediated superconductivity in electron-doped layered TMDC materials, observed using ionic liquid gating technology.
Furthermore, the discovered spin-splitting at the Fermi level could offer crucial experimental validation for theoretical models of Ising-type superconductivity in these materials.
\end{abstract}

\maketitle
\section{Introduction}
Polarons are collective excitation of electrons interacting with the lattice.
These quasiparticles describe an electron moving in a crystal with positively charged ions displacing from their equilibrium positions to effectively screen negative electron charge~\cite{Landau_PZS_1933,Landau_Pekar_ZETPh_1948}.
This coupling between the electron and lattice vibrations (phonon cloud) lowers the electron mobility and increases the electron's effective mass.

The resulting polaronic bound state can be described by Fr\"{o}hlich or Holstein Hamiltonians.
The Fr\"{o}hlich model describes the interaction of electrons with phonons, which leads to the formation of large polarons, whose radii are much larger than the lattice constant. 
The Holstein model, on the other hand, describes the short-range interaction with phonons, which typically leads to the formation of small polarons, whose radii are of the same order of magnitude as the lattice constant~\cite{Emin_2013_polarons}.

Formation of the polarons due to a momentum dependent electron-phonon coupling leads to new features in the electronic spectral function at specific electron valleys, e.g. the emergence of discrete energy levels and discontinuities in the density of states~\cite{Franchini_NRM_2021_polarons}.
In a recent study of alkali-metal dosing on MoS$_2$ surface such discontinuities in density of states have been experimentally observed at the ${\rm K}$ points of the Brillouin zone using angle-resolved photoemission spectroscopy (ARPES)~\cite{Kang_NatMat2018}.
Understanding the changing nature of the quasiparticles as a function of doping is an important prerequisite for understanding the onset of superconductivity observed in highly electron-doped MoS2 under liquid ionic gating~\cite{Ye_Science_2012, Lu_Science_2015}.

Theoretical considerations point to two key novelties in the case of multi-valley electron-doped TDMCs.
First, polaron modes should be observed around both ${\rm K}$ and $\Sigma$ valleys in MoS$_2$~\cite{Garcia_CommPhys_2019}.
\begin{figure}[hbt!]     
\centering
\includegraphics[width=0.48\textwidth]{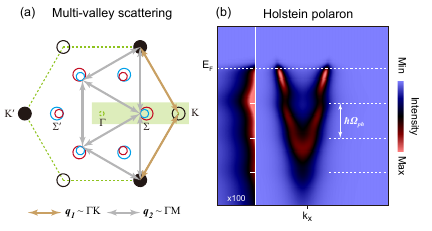}
\caption{\textbf{Phonon-mediated polarons in electron doped MoSe$_2$:}
(a) Multi-valley scattering of conduction electrons in the first Brillouin zone of MoSe$_2$;
(b) Electron-like parabolic band at ${\rm K}$-valley with Holstein polarons simulated by momentum-average model~\cite{Goodvin_PRB_2006}.}
\label{Fig_1}
\end{figure}
Figure~\ref{Fig_1}(a) shows the multi-valley scattering between $\Sigma$ and ${\rm K}$ valleys in the Brillouin zone. 
The electron scattering vectors $q_1$ and $q_2$ correspond to either $\Gamma$-${\rm K}$ or $\Gamma$-${\rm M}$ phonon wave vectors. 
The dominant contribution to electron-phonon coupling arises from the highest energy acoustic phonon mode. 
This behaviour has been demonstrated both experimentally~\cite{Zeng_NatureNanotech_2012} and theoretically~\cite{Kaasbjerg_PRB_2012} for MoS$_2$. 
As in the case of MoS$_2$~\cite{Kaasbjerg_PRB_2012, Zhao_NJP_2018}, for MoSe$_2$ the acoustic phonon modes have essentially flat dispersion over an extended region of $k$-space between the ${\rm M}$ and ${\rm K}$ points~\cite{Bae_NatCommun_2022}. 
Thus, despite their acoustic nature, these phonon modes with characteristic frequency $\Omega_{ph}$ define the energy scale for electron-phonon coupling and corresponding polaron formation~\cite{Zhao_NJP_2018, Bae_NatCommun_2022}.
However, only ${\rm K}$ valley Holstein polarons (Fig.~\ref{Fig_1}(b)) have been observed in the ARPES study of Rb-dosed MoS$_2$ surface ~\cite{Kang_NatMat2018}. 
Second, the spin-orbit coupling (SOC) should induce the splitting of the electron bands at $\Sigma$ and $\Sigma$' valleys due to inversion symmetry breaking on the surface with alkali metal dosing. 
However, in all previous ARPES studies of alkali-dosed semiconducting TMDCs the spin-split features at $\Sigma$ valley have not been resolved~\cite{Zhang_NatNanoTech2014, Eknapakul_NanoLett2014, Alidoust_NatComm2014, Riley_NatNanoTech2015, Miwa_PRL2015, Zhang_NanoLett2016, Kim_SciRep2016, Kim_SciRep2017, Kang_NanoLett2017, Kang_NatMat2018, Han_PRL2021}. 
\begin{figure*}[!t]     
\centering
\includegraphics[width=1\textwidth]{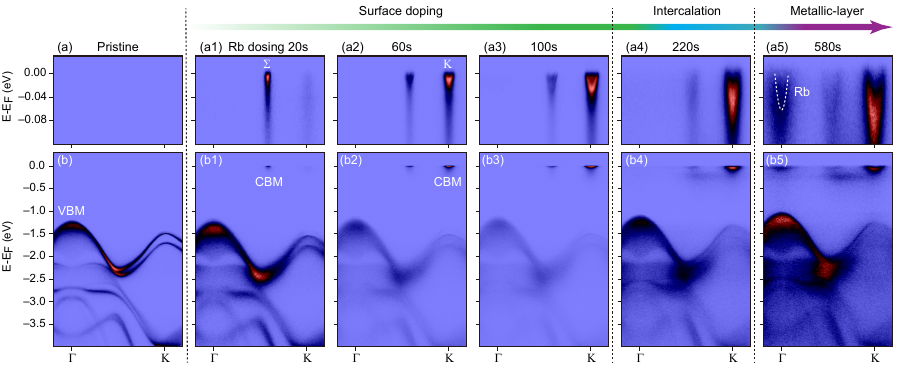}
\caption{\textbf{Evolution of the MoSe$_2$ electronic structure with Rb dosing:}
(a-a5) ARPES spectra of $\Gamma$-${\rm K}$ conduction band dispersions measured near the Fermi energy for cumulative dosing time 0, 20, 60, 100, 220, 580s.  
(b-b5) Corresponding valence band spectra in a wide energy range.}
\label{Fig_2}
\end{figure*}

The isostructural MoSe$_2$ compound is the perfect model system for probing the multi-valley physics induced by electron-doping.
Density functional theory (DFT) calculations show that in contrast to MoS$_2$, the conduction band minimum (CBM) in bulk MoSe$_2$ is at the $\Sigma$, rather than at ${\rm K}$ valley (see Supplementary Fig.~\ref{Fig_S_DFT}).
For the alkali-metal dosing on MoSe$_2$ surface one expects to have larger SOC band-splitting than in the case of MoS$_2$. 
The accessibility of the $\Sigma$ valley, as well as the ${\rm K}$ valley expected at higher doping~\cite{Kim_SciRep2016,Kim_SciRep2017}, makes MoSe$_2$ the best candidate to study the physics of multi-valley polaron modes using high-resolution ARPES.

\section{Results and discussion}

\subsection{Evolution of the electronic band structure with dosing}
In our ARPES data shown in Fig.~\ref{Fig_2} the spectral intensity appears at the Fermi level, marking the occupation of the conduction band and corresponding semiconductor to conductor transition with the Rb dosing on the  MoSe$_2$ surface.
After initial dosing the maximum spectral weight at the Fermi level is observed around $\Sigma$ valley, showing an indirect band gap $E_g^{\Gamma-\Sigma}$~$\sim$1.4\,eV.
However, with increasing amount of dosed Rb the relative intensity of the $\Sigma$ valley decreases, whereas the intensity of ${\rm K}$ valley increases. 
Correspondingly, it appears that the conduction band minimum shifts from $\Sigma$ to ${\rm K}$ valley as in the case of monolayer MoSe$_2$ (see Fig.~\ref{Fig_2}(a1-a3)).
Withing observed energy broadening the band minima at $\Sigma$ and ${\rm K}$ are very close (see Supplementary Fig.~\ref{Fig_S_EDM_CB_VB}), incidentally making electron doped MoSe$_2$ a perfect system for studying multi-valley physics.
Simultaneously with the shift of the CBM intensity, the intensity of the valence band compared to the conduction band decreases (see Fig.~\ref{Fig_2} (b1-b3)).
This first crossover at around 40\,s of Rb dosing marks the transition from the electron doped bulk to the more ``monolayer-like" doped surface conduction band structure~\cite{Eknapakul_NanoLett2014, Riley_NatNanoTech2015, Kang_NanoLett2017, Kim_SciRep2017}.

The accumulation of alkali atoms on the surface and increased electron doping of the top-layer causes a corresponding increase of the electric field, and results in the amplification of the inversion symmetry-breaking at the surface.
At dosing above this first crossover the measured electronic structure is a superposition of the bands from both the bulk material and the doped surface.
This observation is supported by our Se 3d core level and ${\rm K}$-valley valence band data (see Supplementary Fig.~\ref{Fig_S_XPS} and Fig.~\ref{Fig_S_EDM_CB_VB} correspondingly).

At Rb dosing times above 120\,s the relative intensity of the valence bands increases due to the interlayer intercalation of Rb atoms~\cite{Eknapakul_NanoLett2014}.
At the same time the energy difference between bulk and surface Se 3d core levels or ``effective chemical shift" saturates and does not increase with Rb deposition time (see Supplementary Fig.~\ref{Fig_S_XPS}).
This corresponds to the second crossover from surface doping to interlayer intercalation regime.
For dosing times above 340\,s one start to observe build up of the photoemission intensity at the Fermi level around $\Gamma$-point as seen in Fig.~\ref{Fig_2}(a5).
This corresponds to the formation of the metallic Rb states on the MoSe$_2$ surface~\cite{Kim_SciRep2017} and the third crossover from interlayer intercalation regime to growth of ordered alkali metal layer on the surface. 

\subsection{Holstein polarons in ``surface doping" regime}
\begin{figure}[t]     
\centering
\includegraphics[width=0.5\textwidth]{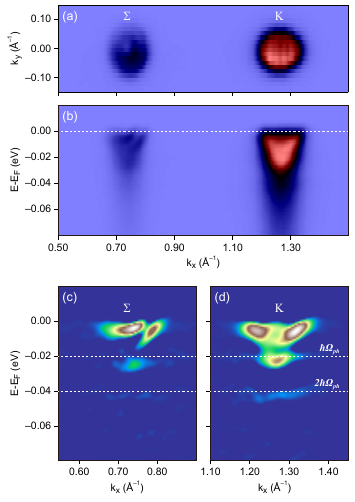}
\caption{\textbf{Formation of polarons in the electron-doped  MoSe$_2$:}
(a) Fermi surface and (b) $\Sigma$-${\rm K}$ band dispersion measured after 120\,s Rb dosing;
Corresponding second derivative of ARPES intensity at $\Sigma$ (c) and ${\rm K}$ (d) valleys.}
\label{Fig_3}
\end{figure}

The most intriguing effects in the electronic structure are observed in  the ``surface doping" regime, corresponding to dosing times between first and second crossover.
Figure~\ref{Fig_3}(a,b) shows the electronic band structure of MoSe$_2$ at Fermi level at 120\,s Rb-dosing on the surface. 
The measured Fermi surface shows two electron-like pockets at $\Sigma$ valley and single electron-like pocket at ${\rm K}$ valley. 
The second derivative images of $\Sigma$ and ${\rm K}$ valleys in Fig.~\ref{Fig_3}(c,d) both show well-defined discontinuities of the spectral function separated by 20\,meV.
Moreover, the photoemission spectra integrated withing a selected momentum range between $\Sigma$ and ${\rm K}$ valleys in Fig.~\ref{Fig_5}(a), show discrete non-dispersing energy levels with the same energy separation (dashed vertical black lines in Fig.~\ref{Fig_5}(b)).

The energy of these spectral features is consistent with the calculated acoustic phonon mode, and their non-dispersing character in an extended momentum space, all point toward formation of small range polarons.%
Therefore, for the electron doped MoSe$_2$ Holstein type polarons are observed at both ${\rm K}$ and $\Sigma$ valleys.

\subsection{Spin-splitting at the $\Sigma$ valley in ``surface doping" regime}
\begin{figure}[tb]     
\centering
\includegraphics[width=0.5\textwidth]{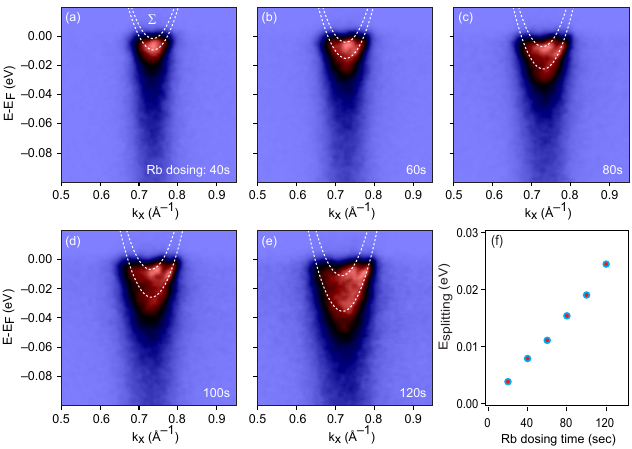}
\caption{\textbf{Evolution of the spin-orbit coupling with Rb dosing:}
(a-e) Band dispersion at $\Sigma$ valley with Rb dosing. 
White doted lines are fits with Rashba-like model with $\Gamma$ been time-reversal invariant momentum (TRIM) point.
(f) Extracted spin-split energy at $\Sigma$ valley.}
\label{Fig_4}
\end{figure}

While polarons at the spin-degenerate ${\rm K}$-valley has been observed previously for MoS$_2$~\cite{Kang_NatMat2018}, polarons at the spin-polarised $\Sigma$ valleys are discovered for the first time.
The spin-orbit splitting of the electron band dispersion at the $\Sigma$ valley is apparent in ARPES data measured along the $\Gamma$-${\rm K}$ direction. 
This agrees with theoretical predictions of lifting the spin degeneracy under electron doping and much larger spin-orbit splitting at $\Sigma$ compared to ${\rm K}$ due to distinct orbital character of conduction bands at these valleys~\cite{Brumme_PRB_2015, Clark_PRB2019, Boccuni_JPCL_2024}.
Figure~\ref{Fig_4} shows band dispersions at the $\Sigma$ valley together with the Rashba-like model fit constrained by the band positions from the corresponding momentum distribution curves (MDCs) at the Fermi level. 
The increase of the Fermi surface size indicates the gradual electron doping of MoSe$_2$ with Rb dosing. 
From the same data one could see that the magnitude of band-splitting is also increasing with electron doping.
This Rashba-like spin-splitting of the bands comes from inversion symmetry breaking due to the electric field created by electron charge transfer from dosed alkali atoms at the surface.
Such inversion symmetry breaking on the surface has been seen at the valence band of semiconducting~\cite{Kang_NanoLett2017, Kim_SciRep2017} as well as at the conduction band of semimetallic~\cite{Clark_PRB2019} alkali dosed TMDC materials.
Figure~\ref{Fig_4}(f) shows how the magnitude of the observed splitting consistently and monotonically becomes larger as the surface charge increases.
This observation aligns perfectly with our understanding that the band splitting is Rashba-like and arises from the combination of spin-orbit coupling with the increasing inversion symmetry-breaking electric field at the surface. 
Moreover, the constancy of intensity for the two components with dosing rules out the case of superposition of the surface and subsurface as a possible origin of the observed splitting.

\subsection{Plasmonic flat states in intercalation regime}
\begin{figure}[t]     
\centering
\includegraphics[width=0.5\textwidth]{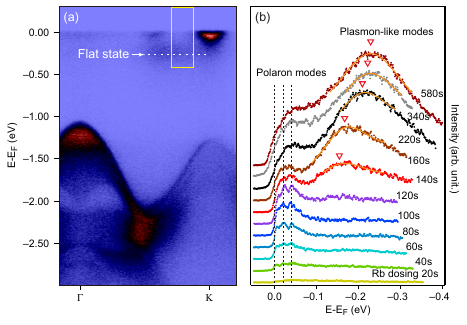}
\caption{\textbf{Crossover from polarons to plasmons in the MoSe$_2$ electronic structure with Rb dosing:}
(a) Wide energy range $\Gamma$-${\rm K}$ band dispersion measured after 340\,s Rb dosing. A white arrow indicates a flat state between $\Sigma$ and ${\rm K}$ valleys; 
(b) Integrated Energy Distribution Curves (EDCs) between $\Sigma$ and ${\rm K}$ valleys (yellow box in (a)).}
\label{Fig_5}
\end{figure}

At dosing times above 340\,s one observes a formation of the flat states in the vicinity of the CBM with much higher binding energies than for Holstein polaronic modes (white arrow at Fig.~\ref{Fig_5}(a)).
To emphasize the difference between these new flat states and Holstein polarons, we analysed momentum integrated EDCs between $\Sigma$ and ${\rm K}$ valleys in Fig.\ref{Fig_5}(a). 
Fig.~\ref{Fig_5}(b) shows that indeed two different features are visible in the integrated EDCs. 
The characteristic binding energies of the polaronic states in the ``surface doping" regime are constant and correspond to the acoustic phonon frequency $\simeq$20\,meV, but the higher binding energies of the flat states in the ``interlayer intercalation" regime significantly increase with dosing.
While binding energies of these flat states are an order of magnitude larger that of Holstein polarons, they are also an order of magnitude smaller then the 3.4\,eV bulk plasmon frequency of metallic Rb~\cite{vomFelde_EPL_1987}.
Binding energies of the flat states as a function 2D electron carrier density can be fitted with plasmon-like dispersion $\hbar\omega_{p} = \sqrt{(n_{e}^{\Sigma} + n_{e}^{K}) e^{2}/ m \epsilon}$ .
The obtained dielectric constant $\epsilon$$\simeq$14.3 is comparable with the calculated in-plane dielectric constant for monolayer MoSe$_2$~\cite{Laturia_2DMatt_2018}.
This suggests that these flat states may indeed be plasmonic polarons, similar to the case of EuO~\cite{Riley_NatCommun_2018} and anatase TiO$_2$~\cite{Ma_NanoLett_2020}. 
Plasmonic polarons have been observed by ARPES in electron doped MoS$_2$~\cite{Caruso_PRB2021}, MoS$_2$/TiO$_2$~\cite{Xiang_JPCL_2023}, HfS$_2$~\cite{Emeis_PRB2023} and WS$_2$~\cite{Ulstrup_arxiv_2308_16509}.
%
\section{Conclusion}
\begin{figure}[tb!]     
\centering
\includegraphics[width=0.5\textwidth]{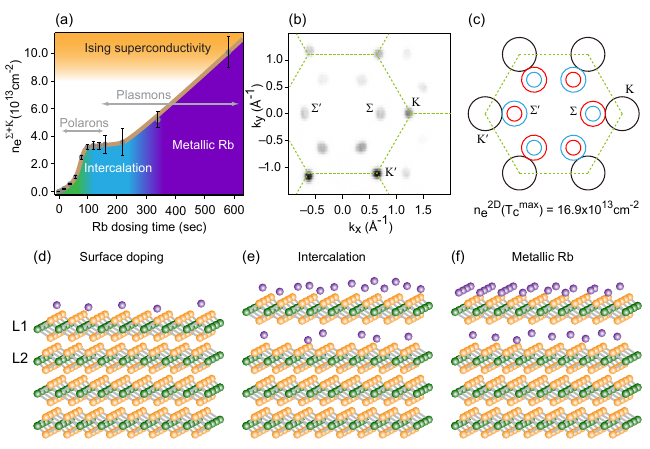}
\caption{\textbf{Phase diagram and Fermi surface of MoSe$_2$ with electron doping:}
(a) 2D electron carrier density as a function of the Rb-dosing time;
(b) Measured Fermi surface of the electron doped MoSe$_2$ after 140\,s Rb dosing. 
(c) Fermi surface extrapolated for maximum T$_{\rm c}$~$\sim$7\,K carrier density doping level~\cite{Shi_SciRep_2015};
(d,e, f) Schematic of the crystal structure for surface dosing, intercalation and metallic Rb regimes. L1 and L2 notes topmost and second topmost layers, correspondingly.
}
\label{Fig_6}
\end{figure}

By studying electronic structure evolution of MoSe$_2$ with Rb surface dosing we have identified three different regimes: surface doping, inter-layer intercalation and ordered alkali metal layer formation.
Figure \ref{Fig_6}(a) shows the two dimensional (2D) electron carrier density as a function of the alkali metal dosing time, calculated using the Luttinger's theorem and experimentally determined sizes of the Fermi surface.
In the ``surface doping" regime we observe continuous electron doping of MoSe$_2$ with corresponding increase of the Fermi surface size and value of the Rashba-like spin-orbit splitting. 
In this regime, before interlayer intercalation occurs, Rb-dosing on the surface is comparable to doping induced by the electrostatic gating using ionic liquid. 
Therefore, the electronic structure on MoSe$_2$ under Rb dosing in ``surface doping" regime corresponds to the electronic structure of MoSe$_2$ in ionic liquid gating device under positive gate voltage (electron doping).
%

The changing nature of the quasiparticles at the Fermi level due to electron-phonon coupling, as well as the evolution of the Fermi surface topology and the electronic density of states, plays a crucial role in understanding superconductivity in electron-doped TMDC semiconductors discovered using liquid gating devices~\cite{Taniguchi_APL_2012, Ye_Science_2012, Shi_SciRep_2015, Lu_Science_2015}.
In the case of MoSe$_2$ the superconductivity occurs at high 2D electron carrier densities $\geq$8$\cdot$10$^{13}$\,cm$^{-2}$ and maximum T$_{\rm c}$~$\sim$7\,K is achieved at electron density $\simeq$16.9$\cdot$10$^{13}$\,cm$^{-2}$, correspondingly~\cite{Shi_SciRep_2015}.

While our results show that the increasing surface electric field description is no longer applicable when we dose Rb atoms above 120\,s, we could extrapolate a gradual increase of the 2D electron carrier density with the alkali metal dosing time within the "surface doping" regime up to the superconducting transition. 
In this case, using extrapolation from our ARPES data we conclude that maximum T$_{\rm c}$ doping level corresponds to the Fermi surface topology where the electron-like pocket at ${\rm K}$ valley touches the outer electron pocket at $\Sigma$ valley (see Fig.~\ref{Fig_6}(b)).
In another words, the maximum T$_{\rm c}$ corresponds to the Fermi level being located exactly at the conduction band saddle point between ${\rm K}$ and $\Sigma$ valleys. 
This can lead to a dramatic enhancement of pairing strength, due to strong increase of the momentum resolved electron-phonon coupling constant $\lambda$, like in case of Rb intercalated PtTe$_2$ ~\cite{Wu_PRB_2021}.
It can be assumed that with a further increase in doping, the electronic structure will undergo a Lifshitz transition, significantly reducing the density of states at the Fermi level and, correspondingly, suppressing electronic pairing.
This can naturally explain the observed superconducting dome for T$_{\rm c}$ vs. doping phase diagram~\cite{Ye_Science_2012, Shi_SciRep_2015}.

Using ionic liquid gating devices, it was found that for electric-field-induced superconductivity in MoS$_2$ the observed upper critical field is four-five times exceeds the Pauli limit~\cite{Lu_Science_2015, Saito_NaturePhys_2016}.
This violation of the paramagnetic limit and significant spin-orbit coupling at the Fermi level suggest a non-spin-singlet pairing that is robust against an external magnetic field.
Ising pairing for the valley alternating spin-split Fermi surface has been suggested in a several theoretical works for non-centrosymmetric TMDC superconductors~\cite{Zhou_PRB_2016, Ilic_PRL_2017, Wang_PRL_2019, Liu_PRB_2020, Wickramaratne_PRX_2020, Semenov_JETP_Lrtt_2024}.
However, all these models consider basic Fermi surface topology with spin-degenerate states at ${\rm K}$ valleys only. 
At the same time our study shows the direct evidence of Rashba-like splitting of the bands at $\Sigma$ valley for such in-situ electron doped materials.

An important consequence of this discovery is that a significant part of the total Fermi surface will be spin split at the optimal doping.
This agrees with the theoretical prediction~\cite{Zhao_NJP_2020} that with anti-symmetric field effect doping the resulting Fermi surface is dominantly formed by the pockets around $\Sigma$ valleys (only $\sim$30\% of its area is around ${\rm K}$ valleys as shown in Fig.\ref{Fig_6}(b)).
Thus any realistic pairing model of Ising superconductivity in electron doped TMDCs should include such composite spin-split multi-valley Fermi surface topology and corresponding interband scattering processes.

To summarise, we observed, for the first time, spin-polarised Holstein type polaronic quasiparticles on Rb dosed MoSe$_2$ surfaces. 
This is direct evidence of the predicted Rashba-like spin-orbit split polarons~\cite{Covaci_PRL_2009}. 
We found that beyond the ``surface doping" regime threshold, when dosed alkali atoms intercalate into the van der Waals (vdW) gap between the layers, completely new plasmon-like quasiparticles emerge. Our findings not only suggest the crucial role of polaron formation for the electronic structure of electron-doped transition-metal dichalcogenides, but also provide experimental foundation for theoretical models of Ising-type superconductivity in these materials.
%

\section{Methods}

High quality single crystals of MoSe$_2$ were grown by the chemical vapor transport technique \cite{Legma_JCG1993}. 
ARPES measurements were performed at the I05 beamline at the Diamond Light Source, UK~\cite{hoesch2017facility}. 
The photoelectron's energy and angular distributions were analysed with a SCIENTA R4000 hemispherical analyser. 
The angular resolution was 0.2$^\circ$, and the overall energy resolution was better than 5\,meV. 
P-polarised 60\,eV photons corresponding to the $\Gamma$ point of the bulk Brillouin zone were used for high-resolution ARPES, and 120\,eV photons for Se 3d core level measurements.
Single crystal samples were cleaved and measured in ultra high-vacuum at temperatures below 10\,K and pressures below 1$\cdot$10$^{-10}$\,mbar. 
Rb atoms were dosed in several sequences on a cold MoSe$_2$ surface using a commercial SAES alkali-metal dispenser. 
The density functional theory (DFT) calculations including spin-orbit coupling with generalized gradient approximation (GGA) were performed using Wien2K package~\cite{Blaha_JCP_2020wien2k}. 
Spectral function simulation for polarons has been done by using momentum average (MA) approximation~\cite{Goodvin_PRB_2006} with 20\,meV constant phonon energy. 
%

\section{Acknowledgements}
We acknowledge Diamond Light Source for time on Beamline I05-ARPES under Proposal NT26631.
S. W. Jung was supported by the National Research Foundation of Korea (NRF) grant funded by the Korea government (MSIT) (No. NRF-2022R1C1C1004978) and supported by Learning \& Academic research institution for Master’s·PhD students, and Postdocs (LAMP) Program of the National Research Foundation of Korea (NRF) grant funded by the Ministry of Education (No. RS-2023-00301974).

\clearpage
\bibliography{MoSe2_polarons_ref_new.bib}

\begin{thebibliography}{50}%
\makeatletter
\providecommand \@ifxundefined [1]{%
 \@ifx{#1\undefined}
}%
\providecommand \@ifnum [1]{%
 \ifnum #1\expandafter \@firstoftwo
 \else \expandafter \@secondoftwo
 \fi
}%
\providecommand \@ifx [1]{%
 \ifx #1\expandafter \@firstoftwo
 \else \expandafter \@secondoftwo
 \fi
}%
\providecommand \natexlab [1]{#1}%
\providecommand \enquote  [1]{``#1''}%
\providecommand \bibnamefont  [1]{#1}%
\providecommand \bibfnamefont [1]{#1}%
\providecommand \citenamefont [1]{#1}%
\providecommand \href@noop [0]{\@secondoftwo}%
\providecommand \href [0]{\begingroup \@sanitize@url \@href}%
\providecommand \@href[1]{\@@startlink{#1}\@@href}%
\providecommand \@@href[1]{\endgroup#1\@@endlink}%
\providecommand \@sanitize@url [0]{\catcode `\\12\catcode `\$12\catcode `\&12\catcode `\#12\catcode `\^12\catcode `\_12\catcode `\%12\relax}%
\providecommand \@@startlink[1]{}%
\providecommand \@@endlink[0]{}%
\providecommand \url  [0]{\begingroup\@sanitize@url \@url }%
\providecommand \@url [1]{\endgroup\@href {#1}{\urlprefix }}%
\providecommand \urlprefix  [0]{URL }%
\providecommand \Eprint [0]{\href }%
\providecommand \doibase [0]{http://dx.doi.org/}%
\providecommand \selectlanguage [0]{\@gobble}%
\providecommand \bibinfo  [0]{\@secondoftwo}%
\providecommand \bibfield  [0]{\@secondoftwo}%
\providecommand \translation [1]{[#1]}%
\providecommand \BibitemOpen [0]{}%
\providecommand \bibitemStop [0]{}%
\providecommand \bibitemNoStop [0]{.\EOS\space}%
\providecommand \EOS [0]{\spacefactor3000\relax}%
\providecommand \BibitemShut  [1]{\csname bibitem#1\endcsname}%
\let\auto@bib@innerbib\@empty
\bibitem [{\citenamefont {Landau}(1933)}]{Landau_PZS_1933}%
  \BibitemOpen
  \bibfield  {author} {\bibinfo {author} {\bibfnamefont {Lev~Davidovich}\ \bibnamefont {Landau}},\ }\bibfield  {title} {\enquote {\bibinfo {title} {Electron motion in crystal lattices},}\ }\href {https://doi.org/10.1016/b978-0-08-010586-4.50015-8} {\bibfield  {journal} {\bibinfo  {journal} {Phys. Z. Sowjet.}\ }\textbf {\bibinfo {volume} {3}},\ \bibinfo {pages} {664} (\bibinfo {year} {1933})}\BibitemShut {NoStop}%
\bibitem [{\citenamefont {Landau}\ and\ \citenamefont {Pekar}(1948)}]{Landau_Pekar_ZETPh_1948}%
  \BibitemOpen
  \bibfield  {author} {\bibinfo {author} {\bibfnamefont {LD}~\bibnamefont {Landau}}\ and\ \bibinfo {author} {\bibfnamefont {SI}~\bibnamefont {Pekar}},\ }\bibfield  {title} {\enquote {\bibinfo {title} {Effective mass of a polaron},}\ }\href {https://doi.org/10.1016/B978-0-08-010586-4.50072-9} {\bibfield  {journal} {\bibinfo  {journal} {Zh. Eksp. Teor. Fiz}\ }\textbf {\bibinfo {volume} {18}},\ \bibinfo {pages} {419--423} (\bibinfo {year} {1948})}\BibitemShut {NoStop}%
\bibitem [{\citenamefont {Emin}(2013)}]{Emin_2013_polarons}%
  \BibitemOpen
  \bibfield  {author} {\bibinfo {author} {\bibfnamefont {David}\ \bibnamefont {Emin}},\ }\href@noop {} {\emph {\bibinfo {title} {Polarons}}}\ (\bibinfo  {publisher} {Cambridge University Press},\ \bibinfo {year} {2013})\BibitemShut {NoStop}%
\bibitem [{\citenamefont {Franchini}\ \emph {et~al.}(2021)\citenamefont {Franchini}, \citenamefont {Reticcioli}, \citenamefont {Setvin},\ and\ \citenamefont {Diebold}}]{Franchini_NRM_2021_polarons}%
  \BibitemOpen
  \bibfield  {author} {\bibinfo {author} {\bibfnamefont {Cesare}\ \bibnamefont {Franchini}}, \bibinfo {author} {\bibfnamefont {Michele}\ \bibnamefont {Reticcioli}}, \bibinfo {author} {\bibfnamefont {Martin}\ \bibnamefont {Setvin}}, \ and\ \bibinfo {author} {\bibfnamefont {Ulrike}\ \bibnamefont {Diebold}},\ }\bibfield  {title} {\enquote {\bibinfo {title} {Polarons in materials},}\ }\href {https://doi.org/10.1038/s41578-021-00289-w} {\bibfield  {journal} {\bibinfo  {journal} {Nature Reviews Materials}\ }\textbf {\bibinfo {volume} {6}},\ \bibinfo {pages} {560--586} (\bibinfo {year} {2021})}\BibitemShut {NoStop}%
\bibitem [{\citenamefont {Kang}\ \emph {et~al.}(2018)\citenamefont {Kang}, \citenamefont {Jung}, \citenamefont {Shin}, \citenamefont {Sohn}, \citenamefont {Ryu}, \citenamefont {Kim}, \citenamefont {Hoesch},\ and\ \citenamefont {Kim}}]{Kang_NatMat2018}%
  \BibitemOpen
  \bibfield  {author} {\bibinfo {author} {\bibfnamefont {Mingu}\ \bibnamefont {Kang}}, \bibinfo {author} {\bibfnamefont {Sung~Won}\ \bibnamefont {Jung}}, \bibinfo {author} {\bibfnamefont {Woo~Jong}\ \bibnamefont {Shin}}, \bibinfo {author} {\bibfnamefont {Yeongsup}\ \bibnamefont {Sohn}}, \bibinfo {author} {\bibfnamefont {Sae~Hee}\ \bibnamefont {Ryu}}, \bibinfo {author} {\bibfnamefont {Timur~K}\ \bibnamefont {Kim}}, \bibinfo {author} {\bibfnamefont {Moritz}\ \bibnamefont {Hoesch}}, \ and\ \bibinfo {author} {\bibfnamefont {Keun~Su}\ \bibnamefont {Kim}},\ }\bibfield  {title} {\enquote {\bibinfo {title} {Holstein polaron in a valley-degenerate two-dimensional semiconductor},}\ }\href {\doibase https://doi.org/10.1038/s41563-018-0092-7} {\bibfield  {journal} {\bibinfo  {journal} {Nature Materials}\ }\textbf {\bibinfo {volume} {17}},\ \bibinfo {pages} {676--680} (\bibinfo {year} {2018})}\BibitemShut {NoStop}%
\bibitem [{\citenamefont {Ye}\ \emph {et~al.}(2012)\citenamefont {Ye}, \citenamefont {Zhang}, \citenamefont {Akashi}, \citenamefont {Bahramy}, \citenamefont {Arita},\ and\ \citenamefont {Iwasa}}]{Ye_Science_2012}%
  \BibitemOpen
  \bibfield  {author} {\bibinfo {author} {\bibfnamefont {JT}~\bibnamefont {Ye}}, \bibinfo {author} {\bibfnamefont {Yj~J}\ \bibnamefont {Zhang}}, \bibinfo {author} {\bibfnamefont {R}~\bibnamefont {Akashi}}, \bibinfo {author} {\bibfnamefont {Ms~S}\ \bibnamefont {Bahramy}}, \bibinfo {author} {\bibfnamefont {R}~\bibnamefont {Arita}}, \ and\ \bibinfo {author} {\bibfnamefont {Y}~\bibnamefont {Iwasa}},\ }\bibfield  {title} {\enquote {\bibinfo {title} {Superconducting dome in a gate-tuned band insulator},}\ }\href {https://doi.org/10.1126/science.1228006} {\bibfield  {journal} {\bibinfo  {journal} {Science}\ }\textbf {\bibinfo {volume} {338}},\ \bibinfo {pages} {1193--1196} (\bibinfo {year} {2012})}\BibitemShut {NoStop}%
\bibitem [{\citenamefont {Lu}\ \emph {et~al.}(2015)\citenamefont {Lu}, \citenamefont {Zheliuk}, \citenamefont {Leermakers}, \citenamefont {Yuan}, \citenamefont {Zeitler}, \citenamefont {Law},\ and\ \citenamefont {Ye}}]{Lu_Science_2015}%
  \BibitemOpen
  \bibfield  {author} {\bibinfo {author} {\bibfnamefont {JM}~\bibnamefont {Lu}}, \bibinfo {author} {\bibfnamefont {O}~\bibnamefont {Zheliuk}}, \bibinfo {author} {\bibfnamefont {Inge}\ \bibnamefont {Leermakers}}, \bibinfo {author} {\bibfnamefont {Noah~FQ}\ \bibnamefont {Yuan}}, \bibinfo {author} {\bibfnamefont {Uli}\ \bibnamefont {Zeitler}}, \bibinfo {author} {\bibfnamefont {Kam~Tuen}\ \bibnamefont {Law}}, \ and\ \bibinfo {author} {\bibfnamefont {JT}~\bibnamefont {Ye}},\ }\bibfield  {title} {\enquote {\bibinfo {title} {Evidence for two-dimensional ising superconductivity in gated {MoS$_{2}$}},}\ }\href {https://doi.org/10.1126/science.aab2277} {\bibfield  {journal} {\bibinfo  {journal} {Science}\ }\textbf {\bibinfo {volume} {350}},\ \bibinfo {pages} {1353--1357} (\bibinfo {year} {2015})}\BibitemShut {NoStop}%
\bibitem [{\citenamefont {Garcia-Goiricelaya}\ \emph {et~al.}(2019)\citenamefont {Garcia-Goiricelaya}, \citenamefont {Lafuente-Bartolome}, \citenamefont {Gurtubay},\ and\ \citenamefont {Eiguren}}]{Garcia_CommPhys_2019}%
  \BibitemOpen
  \bibfield  {author} {\bibinfo {author} {\bibfnamefont {Peio}\ \bibnamefont {Garcia-Goiricelaya}}, \bibinfo {author} {\bibfnamefont {Jon}\ \bibnamefont {Lafuente-Bartolome}}, \bibinfo {author} {\bibfnamefont {Idoia~G}\ \bibnamefont {Gurtubay}}, \ and\ \bibinfo {author} {\bibfnamefont {Asier}\ \bibnamefont {Eiguren}},\ }\bibfield  {title} {\enquote {\bibinfo {title} {Long-living carriers in a strong electron--phonon interacting two-dimensional doped semiconductor},}\ }\href {https://doi.org/10.1038/s42005-019-0182-0} {\bibfield  {journal} {\bibinfo  {journal} {Communications Physics}\ }\textbf {\bibinfo {volume} {2}},\ \bibinfo {pages} {81} (\bibinfo {year} {2019})}\BibitemShut {NoStop}%
\bibitem [{\citenamefont {Goodvin}\ \emph {et~al.}(2006)\citenamefont {Goodvin}, \citenamefont {Berciu},\ and\ \citenamefont {Sawatzky}}]{Goodvin_PRB_2006}%
  \BibitemOpen
  \bibfield  {author} {\bibinfo {author} {\bibfnamefont {Glen~L}\ \bibnamefont {Goodvin}}, \bibinfo {author} {\bibfnamefont {Mona}\ \bibnamefont {Berciu}}, \ and\ \bibinfo {author} {\bibfnamefont {George~A}\ \bibnamefont {Sawatzky}},\ }\bibfield  {title} {\enquote {\bibinfo {title} {{Green’s function of the Holstein polaron}},}\ }\href {https://doi.org/10.1103/PhysRevB.74.245104} {\bibfield  {journal} {\bibinfo  {journal} {Physical Review B}\ }\textbf {\bibinfo {volume} {74}},\ \bibinfo {pages} {245104} (\bibinfo {year} {2006})}\BibitemShut {NoStop}%
\bibitem [{\citenamefont {Zeng}\ \emph {et~al.}(2012)\citenamefont {Zeng}, \citenamefont {Dai}, \citenamefont {Yao}, \citenamefont {Xiao},\ and\ \citenamefont {Cui}}]{Zeng_NatureNanotech_2012}%
  \BibitemOpen
  \bibfield  {author} {\bibinfo {author} {\bibfnamefont {Hualing}\ \bibnamefont {Zeng}}, \bibinfo {author} {\bibfnamefont {Junfeng}\ \bibnamefont {Dai}}, \bibinfo {author} {\bibfnamefont {Wang}\ \bibnamefont {Yao}}, \bibinfo {author} {\bibfnamefont {Di}~\bibnamefont {Xiao}}, \ and\ \bibinfo {author} {\bibfnamefont {Xiaodong}\ \bibnamefont {Cui}},\ }\bibfield  {title} {\enquote {\bibinfo {title} {Valley polarization in {MoS$_{2}$} monolayers by optical pumping},}\ }\href {\doibase https://doi.org/10.1038/nnano.2012.95} {\bibfield  {journal} {\bibinfo  {journal} {Nature nanotechnology}\ }\textbf {\bibinfo {volume} {7}},\ \bibinfo {pages} {490--493} (\bibinfo {year} {2012})}\BibitemShut {NoStop}%
\bibitem [{\citenamefont {Kaasbjerg}\ \emph {et~al.}(2012)\citenamefont {Kaasbjerg}, \citenamefont {Thygesen},\ and\ \citenamefont {Jacobsen}}]{Kaasbjerg_PRB_2012}%
  \BibitemOpen
  \bibfield  {author} {\bibinfo {author} {\bibfnamefont {Kristen}\ \bibnamefont {Kaasbjerg}}, \bibinfo {author} {\bibfnamefont {Kristian~S.}\ \bibnamefont {Thygesen}}, \ and\ \bibinfo {author} {\bibfnamefont {Karsten~W.}\ \bibnamefont {Jacobsen}},\ }\bibfield  {title} {\enquote {\bibinfo {title} {Phonon-limited mobility in $n$-type single-layer {MoS$_{2}$} from first principles},}\ }\href {\doibase 10.1103/PhysRevB.85.115317} {\bibfield  {journal} {\bibinfo  {journal} {Phys. Rev. B}\ }\textbf {\bibinfo {volume} {85}},\ \bibinfo {pages} {115317} (\bibinfo {year} {2012})}\BibitemShut {NoStop}%
\bibitem [{\citenamefont {Zhao}\ \emph {et~al.}(2018)\citenamefont {Zhao}, \citenamefont {Dai}, \citenamefont {Zhang}, \citenamefont {Lian}, \citenamefont {Zeng}, \citenamefont {Li}, \citenamefont {Meng},\ and\ \citenamefont {Ni}}]{Zhao_NJP_2018}%
  \BibitemOpen
  \bibfield  {author} {\bibinfo {author} {\bibfnamefont {Yinchang}\ \bibnamefont {Zhao}}, \bibinfo {author} {\bibfnamefont {Zhenhong}\ \bibnamefont {Dai}}, \bibinfo {author} {\bibfnamefont {Chao}\ \bibnamefont {Zhang}}, \bibinfo {author} {\bibfnamefont {Chao}\ \bibnamefont {Lian}}, \bibinfo {author} {\bibfnamefont {Shuming}\ \bibnamefont {Zeng}}, \bibinfo {author} {\bibfnamefont {Geng}\ \bibnamefont {Li}}, \bibinfo {author} {\bibfnamefont {Sheng}\ \bibnamefont {Meng}}, \ and\ \bibinfo {author} {\bibfnamefont {Jun}\ \bibnamefont {Ni}},\ }\bibfield  {title} {\enquote {\bibinfo {title} {Intrinsic electronic transport and thermoelectric power factor in n-type doped monolayer mos2},}\ }\href@noop {} {\bibfield  {journal} {\bibinfo  {journal} {New Journal of Physics}\ }\textbf {\bibinfo {volume} {20}},\ \bibinfo {pages} {043009} (\bibinfo {year} {2018})}\BibitemShut {NoStop}%
\bibitem [{\citenamefont {Bae}\ \emph {et~al.}(2022)\citenamefont {Bae}, \citenamefont {Matsumoto}, \citenamefont {Raebiger}, \citenamefont {Shudo}, \citenamefont {Kim}, \citenamefont {Handeg{\aa}rd}, \citenamefont {Nagao}, \citenamefont {Kitajima}, \citenamefont {Sakai}, \citenamefont {Zhang} \emph {et~al.}}]{Bae_NatCommun_2022}%
  \BibitemOpen
  \bibfield  {author} {\bibinfo {author} {\bibfnamefont {Soungmin}\ \bibnamefont {Bae}}, \bibinfo {author} {\bibfnamefont {Kana}\ \bibnamefont {Matsumoto}}, \bibinfo {author} {\bibfnamefont {Hannes}\ \bibnamefont {Raebiger}}, \bibinfo {author} {\bibfnamefont {Ken-ichi}\ \bibnamefont {Shudo}}, \bibinfo {author} {\bibfnamefont {Yong-Hoon}\ \bibnamefont {Kim}}, \bibinfo {author} {\bibfnamefont {{\O}rjan~Sele}\ \bibnamefont {Handeg{\aa}rd}}, \bibinfo {author} {\bibfnamefont {Tadaaki}\ \bibnamefont {Nagao}}, \bibinfo {author} {\bibfnamefont {Masahiro}\ \bibnamefont {Kitajima}}, \bibinfo {author} {\bibfnamefont {Yuji}\ \bibnamefont {Sakai}}, \bibinfo {author} {\bibfnamefont {Xiang}\ \bibnamefont {Zhang}},  \emph {et~al.},\ }\bibfield  {title} {\enquote {\bibinfo {title} {K-point longitudinal acoustic phonons are responsible for ultrafast intervalley scattering in monolayer mose2},}\ }\href@noop {} {\bibfield  {journal} {\bibinfo  {journal} {Nature Communications}\ }\textbf {\bibinfo {volume} {13}},\ \bibinfo
  {pages} {4279} (\bibinfo {year} {2022})}\BibitemShut {NoStop}%
\bibitem [{\citenamefont {Zhang}\ \emph {et~al.}(2014)\citenamefont {Zhang}, \citenamefont {Chang}, \citenamefont {Zhou}, \citenamefont {Cui}, \citenamefont {Yan}, \citenamefont {Liu}, \citenamefont {Schmitt}, \citenamefont {Lee}, \citenamefont {Moore}, \citenamefont {Chen} \emph {et~al.}}]{Zhang_NatNanoTech2014}%
  \BibitemOpen
  \bibfield  {author} {\bibinfo {author} {\bibfnamefont {Yi}~\bibnamefont {Zhang}}, \bibinfo {author} {\bibfnamefont {Tay-Rong}\ \bibnamefont {Chang}}, \bibinfo {author} {\bibfnamefont {Bo}~\bibnamefont {Zhou}}, \bibinfo {author} {\bibfnamefont {Yong-Tao}\ \bibnamefont {Cui}}, \bibinfo {author} {\bibfnamefont {Hao}\ \bibnamefont {Yan}}, \bibinfo {author} {\bibfnamefont {Zhongkai}\ \bibnamefont {Liu}}, \bibinfo {author} {\bibfnamefont {Felix}\ \bibnamefont {Schmitt}}, \bibinfo {author} {\bibfnamefont {James}\ \bibnamefont {Lee}}, \bibinfo {author} {\bibfnamefont {Rob}\ \bibnamefont {Moore}}, \bibinfo {author} {\bibfnamefont {Yulin}\ \bibnamefont {Chen}},  \emph {et~al.},\ }\bibfield  {title} {\enquote {\bibinfo {title} {{Direct observation of the transition from indirect to direct bandgap in atomically thin epitaxial MoSe$_2$}},}\ }\href {https://doi.org/10.1038/nnano.2013.277} {\bibfield  {journal} {\bibinfo  {journal} {Nature Nanotechnology}\ }\textbf {\bibinfo {volume} {9}},\ \bibinfo {pages} {111--115}
  (\bibinfo {year} {2014})}\BibitemShut {NoStop}%
\bibitem [{\citenamefont {Eknapakul}\ \emph {et~al.}(2014)\citenamefont {Eknapakul}, \citenamefont {King}, \citenamefont {Asakawa}, \citenamefont {Buaphet}, \citenamefont {He}, \citenamefont {Mo}, \citenamefont {Takagi}, \citenamefont {Shen}, \citenamefont {Baumberger}, \citenamefont {Sasagawa} \emph {et~al.}}]{Eknapakul_NanoLett2014}%
  \BibitemOpen
  \bibfield  {author} {\bibinfo {author} {\bibfnamefont {T}~\bibnamefont {Eknapakul}}, \bibinfo {author} {\bibfnamefont {PDC}\ \bibnamefont {King}}, \bibinfo {author} {\bibfnamefont {M}~\bibnamefont {Asakawa}}, \bibinfo {author} {\bibfnamefont {P}~\bibnamefont {Buaphet}}, \bibinfo {author} {\bibfnamefont {R-H}\ \bibnamefont {He}}, \bibinfo {author} {\bibfnamefont {S-K}\ \bibnamefont {Mo}}, \bibinfo {author} {\bibfnamefont {H}~\bibnamefont {Takagi}}, \bibinfo {author} {\bibfnamefont {KM}~\bibnamefont {Shen}}, \bibinfo {author} {\bibfnamefont {F}~\bibnamefont {Baumberger}}, \bibinfo {author} {\bibfnamefont {T}~\bibnamefont {Sasagawa}},  \emph {et~al.},\ }\bibfield  {title} {\enquote {\bibinfo {title} {{Electronic structure of a quasi-freestanding MoS$_2$ monolayer}},}\ }\href {https://doi.org/10.1021/nl4042824} {\bibfield  {journal} {\bibinfo  {journal} {Nano Letters}\ }\textbf {\bibinfo {volume} {14}},\ \bibinfo {pages} {1312--1316} (\bibinfo {year} {2014})}\BibitemShut {NoStop}%
\bibitem [{\citenamefont {Alidoust}\ \emph {et~al.}(2014)\citenamefont {Alidoust}, \citenamefont {Bian}, \citenamefont {Xu}, \citenamefont {Sankar}, \citenamefont {Neupane}, \citenamefont {Liu}, \citenamefont {Belopolski}, \citenamefont {Qu}, \citenamefont {Denlinger}, \citenamefont {Chou} \emph {et~al.}}]{Alidoust_NatComm2014}%
  \BibitemOpen
  \bibfield  {author} {\bibinfo {author} {\bibfnamefont {Nasser}\ \bibnamefont {Alidoust}}, \bibinfo {author} {\bibfnamefont {Guang}\ \bibnamefont {Bian}}, \bibinfo {author} {\bibfnamefont {Su-Yang}\ \bibnamefont {Xu}}, \bibinfo {author} {\bibfnamefont {Raman}\ \bibnamefont {Sankar}}, \bibinfo {author} {\bibfnamefont {Madhab}\ \bibnamefont {Neupane}}, \bibinfo {author} {\bibfnamefont {Chang}\ \bibnamefont {Liu}}, \bibinfo {author} {\bibfnamefont {Ilya}\ \bibnamefont {Belopolski}}, \bibinfo {author} {\bibfnamefont {Dong-Xia}\ \bibnamefont {Qu}}, \bibinfo {author} {\bibfnamefont {Jonathan~D}\ \bibnamefont {Denlinger}}, \bibinfo {author} {\bibfnamefont {Fang-Cheng}\ \bibnamefont {Chou}},  \emph {et~al.},\ }\bibfield  {title} {\enquote {\bibinfo {title} {{Observation of monolayer valence band spin-orbit effect and induced quantum well states in MoX$_2$}},}\ }\href {https://doi.org/10.1038/ncomms5673} {\bibfield  {journal} {\bibinfo  {journal} {Nature Communications}\ }\textbf {\bibinfo {volume} {5}},\ \bibinfo
  {pages} {4673} (\bibinfo {year} {2014})}\BibitemShut {NoStop}%
\bibitem [{\citenamefont {Riley}\ \emph {et~al.}(2015)\citenamefont {Riley}, \citenamefont {Meevasana}, \citenamefont {Bawden}, \citenamefont {Asakawa}, \citenamefont {Takayama}, \citenamefont {Eknapakul}, \citenamefont {Kim}, \citenamefont {Hoesch}, \citenamefont {Mo}, \citenamefont {Takagi} \emph {et~al.}}]{Riley_NatNanoTech2015}%
  \BibitemOpen
  \bibfield  {author} {\bibinfo {author} {\bibfnamefont {Jonathon~Mark}\ \bibnamefont {Riley}}, \bibinfo {author} {\bibfnamefont {Worawat}\ \bibnamefont {Meevasana}}, \bibinfo {author} {\bibfnamefont {Lewis}\ \bibnamefont {Bawden}}, \bibinfo {author} {\bibfnamefont {M}~\bibnamefont {Asakawa}}, \bibinfo {author} {\bibfnamefont {T}~\bibnamefont {Takayama}}, \bibinfo {author} {\bibfnamefont {T}~\bibnamefont {Eknapakul}}, \bibinfo {author} {\bibfnamefont {TK}~\bibnamefont {Kim}}, \bibinfo {author} {\bibfnamefont {M}~\bibnamefont {Hoesch}}, \bibinfo {author} {\bibfnamefont {S-K}\ \bibnamefont {Mo}}, \bibinfo {author} {\bibfnamefont {H}~\bibnamefont {Takagi}},  \emph {et~al.},\ }\bibfield  {title} {\enquote {\bibinfo {title} {Negative electronic compressibility and tunable spin splitting in {WSe$_2$}},}\ }\href {https://doi.org/10.1038/nnano.2015.217} {\bibfield  {journal} {\bibinfo  {journal} {Nature Nanotechnology}\ }\textbf {\bibinfo {volume} {10}},\ \bibinfo {pages} {1043--1047} (\bibinfo {year}
  {2015})}\BibitemShut {NoStop}%
\bibitem [{\citenamefont {Miwa}\ \emph {et~al.}(2015)\citenamefont {Miwa}, \citenamefont {Ulstrup}, \citenamefont {S{\o}rensen}, \citenamefont {Dendzik}, \citenamefont {{\v{C}}abo}, \citenamefont {Bianchi}, \citenamefont {Lauritsen},\ and\ \citenamefont {Hofmann}}]{Miwa_PRL2015}%
  \BibitemOpen
  \bibfield  {author} {\bibinfo {author} {\bibfnamefont {Jill~A}\ \bibnamefont {Miwa}}, \bibinfo {author} {\bibfnamefont {S{\o}ren}\ \bibnamefont {Ulstrup}}, \bibinfo {author} {\bibfnamefont {Signe~G}\ \bibnamefont {S{\o}rensen}}, \bibinfo {author} {\bibfnamefont {Maciej}\ \bibnamefont {Dendzik}}, \bibinfo {author} {\bibfnamefont {Antonija~Grubi{\v{s}}i{\'c}}\ \bibnamefont {{\v{C}}abo}}, \bibinfo {author} {\bibfnamefont {Marco}\ \bibnamefont {Bianchi}}, \bibinfo {author} {\bibfnamefont {Jeppe~Vang}\ \bibnamefont {Lauritsen}}, \ and\ \bibinfo {author} {\bibfnamefont {Philip}\ \bibnamefont {Hofmann}},\ }\bibfield  {title} {\enquote {\bibinfo {title} {Electronic structure of epitaxial single-layer {MoS$_2$}},}\ }\href {http://dx.doi.org/10.1103/PhysRevLett.114.046802} {\bibfield  {journal} {\bibinfo  {journal} {Physical Review Letters}\ }\textbf {\bibinfo {volume} {114}},\ \bibinfo {pages} {046802} (\bibinfo {year} {2015})}\BibitemShut {NoStop}%
\bibitem [{\citenamefont {Zhang}\ \emph {et~al.}(2016)\citenamefont {Zhang}, \citenamefont {Ugeda}, \citenamefont {Jin}, \citenamefont {Shi}, \citenamefont {Bradley}, \citenamefont {Mart{\'\i}n-Recio}, \citenamefont {Ryu}, \citenamefont {Kim}, \citenamefont {Tang}, \citenamefont {Kim} \emph {et~al.}}]{Zhang_NanoLett2016}%
  \BibitemOpen
  \bibfield  {author} {\bibinfo {author} {\bibfnamefont {Yi}~\bibnamefont {Zhang}}, \bibinfo {author} {\bibfnamefont {Miguel~M}\ \bibnamefont {Ugeda}}, \bibinfo {author} {\bibfnamefont {Chenhao}\ \bibnamefont {Jin}}, \bibinfo {author} {\bibfnamefont {Su-Fei}\ \bibnamefont {Shi}}, \bibinfo {author} {\bibfnamefont {Aaron~J}\ \bibnamefont {Bradley}}, \bibinfo {author} {\bibfnamefont {Ana}\ \bibnamefont {Mart{\'\i}n-Recio}}, \bibinfo {author} {\bibfnamefont {Hyejin}\ \bibnamefont {Ryu}}, \bibinfo {author} {\bibfnamefont {Jonghwan}\ \bibnamefont {Kim}}, \bibinfo {author} {\bibfnamefont {Shujie}\ \bibnamefont {Tang}}, \bibinfo {author} {\bibfnamefont {Yeongkwan}\ \bibnamefont {Kim}},  \emph {et~al.},\ }\bibfield  {title} {\enquote {\bibinfo {title} {Electronic structure, surface doping, and optical response in epitaxial {WSe$_2$} thin films},}\ }\href {http://dx.doi.org/10.1021/acs.nanolett.6b00059} {\bibfield  {journal} {\bibinfo  {journal} {Nano Letters}\ }\textbf {\bibinfo {volume} {16}},\ \bibinfo {pages}
  {2485--2491} (\bibinfo {year} {2016})}\BibitemShut {NoStop}%
\bibitem [{\citenamefont {Kim}\ \emph {et~al.}(2016)\citenamefont {Kim}, \citenamefont {Rhim}, \citenamefont {Kim}, \citenamefont {Kim},\ and\ \citenamefont {Park}}]{Kim_SciRep2016}%
  \BibitemOpen
  \bibfield  {author} {\bibinfo {author} {\bibfnamefont {Beom~Seo}\ \bibnamefont {Kim}}, \bibinfo {author} {\bibfnamefont {Jun-Won}\ \bibnamefont {Rhim}}, \bibinfo {author} {\bibfnamefont {Beomyoung}\ \bibnamefont {Kim}}, \bibinfo {author} {\bibfnamefont {Changyoung}\ \bibnamefont {Kim}}, \ and\ \bibinfo {author} {\bibfnamefont {Seung~Ryong}\ \bibnamefont {Park}},\ }\bibfield  {title} {\enquote {\bibinfo {title} {{Determination of the band parameters of bulk 2H-MX$_2$ (M= Mo, W; X= S, Se) by angle-resolved photoemission spectroscopy}},}\ }\href {https://doi.org/10.1038/srep36389} {\bibfield  {journal} {\bibinfo  {journal} {Scientific reports}\ }\textbf {\bibinfo {volume} {6}},\ \bibinfo {pages} {36389} (\bibinfo {year} {2016})}\BibitemShut {NoStop}%
\bibitem [{\citenamefont {Kim}\ \emph {et~al.}(2017)\citenamefont {Kim}, \citenamefont {Kyung}, \citenamefont {Seo}, \citenamefont {Kwon}, \citenamefont {Denlinger}, \citenamefont {Kim},\ and\ \citenamefont {Park}}]{Kim_SciRep2017}%
  \BibitemOpen
  \bibfield  {author} {\bibinfo {author} {\bibfnamefont {Beom~Seo}\ \bibnamefont {Kim}}, \bibinfo {author} {\bibfnamefont {WS}~\bibnamefont {Kyung}}, \bibinfo {author} {\bibfnamefont {JJ}~\bibnamefont {Seo}}, \bibinfo {author} {\bibfnamefont {JY}~\bibnamefont {Kwon}}, \bibinfo {author} {\bibfnamefont {JD}~\bibnamefont {Denlinger}}, \bibinfo {author} {\bibfnamefont {Changyoung}\ \bibnamefont {Kim}}, \ and\ \bibinfo {author} {\bibfnamefont {SR}~\bibnamefont {Park}},\ }\bibfield  {title} {\enquote {\bibinfo {title} {Possible electric field induced indirect to direct band gap transition in {MoSe$_{2}$}},}\ }\href {\doibase https://doi.org/10.1038/s41598-017-05613-5} {\bibfield  {journal} {\bibinfo  {journal} {Scientific Reports}\ }\textbf {\bibinfo {volume} {7}},\ \bibinfo {pages} {5206} (\bibinfo {year} {2017})}\BibitemShut {NoStop}%
\bibitem [{\citenamefont {Kang}\ \emph {et~al.}(2017)\citenamefont {Kang}, \citenamefont {Kim}, \citenamefont {Ryu}, \citenamefont {Jung}, \citenamefont {Kim}, \citenamefont {Moreschini}, \citenamefont {Jozwiak}, \citenamefont {Rotenberg}, \citenamefont {Bostwick},\ and\ \citenamefont {Kim}}]{Kang_NanoLett2017}%
  \BibitemOpen
  \bibfield  {author} {\bibinfo {author} {\bibfnamefont {Mingu}\ \bibnamefont {Kang}}, \bibinfo {author} {\bibfnamefont {Beomyoung}\ \bibnamefont {Kim}}, \bibinfo {author} {\bibfnamefont {Sae~Hee}\ \bibnamefont {Ryu}}, \bibinfo {author} {\bibfnamefont {Sung~Won}\ \bibnamefont {Jung}}, \bibinfo {author} {\bibfnamefont {Jimin}\ \bibnamefont {Kim}}, \bibinfo {author} {\bibfnamefont {Luca}\ \bibnamefont {Moreschini}}, \bibinfo {author} {\bibfnamefont {Chris}\ \bibnamefont {Jozwiak}}, \bibinfo {author} {\bibfnamefont {Eli}\ \bibnamefont {Rotenberg}}, \bibinfo {author} {\bibfnamefont {Aaron}\ \bibnamefont {Bostwick}}, \ and\ \bibinfo {author} {\bibfnamefont {Keun~Su}\ \bibnamefont {Kim}},\ }\bibfield  {title} {\enquote {\bibinfo {title} {Universal mechanism of band-gap engineering in transition-metal dichalcogenides},}\ }\href {http://dx.doi.org/10.1021/acs.nanolett.6b04775} {\bibfield  {journal} {\bibinfo  {journal} {Nano Letters}\ }\textbf {\bibinfo {volume} {17}},\ \bibinfo {pages} {1610--1615} (\bibinfo {year}
  {2017})}\BibitemShut {NoStop}%
\bibitem [{\citenamefont {Han}\ \emph {et~al.}(2021)\citenamefont {Han}, \citenamefont {Chen}, \citenamefont {Cai}, \citenamefont {Wang}, \citenamefont {Wang}, \citenamefont {Xin},\ and\ \citenamefont {Zhang}}]{Han_PRL2021}%
  \BibitemOpen
  \bibfield  {author} {\bibinfo {author} {\bibfnamefont {TT}~\bibnamefont {Han}}, \bibinfo {author} {\bibfnamefont {L}~\bibnamefont {Chen}}, \bibinfo {author} {\bibfnamefont {C}~\bibnamefont {Cai}}, \bibinfo {author} {\bibfnamefont {ZG}~\bibnamefont {Wang}}, \bibinfo {author} {\bibfnamefont {YD}~\bibnamefont {Wang}}, \bibinfo {author} {\bibfnamefont {ZM}~\bibnamefont {Xin}}, \ and\ \bibinfo {author} {\bibfnamefont {Y}~\bibnamefont {Zhang}},\ }\bibfield  {title} {\enquote {\bibinfo {title} {{Metal-Insulator Transition and Emergent Gapped Phase in the Surface-Doped 2D Semiconductor 2H-MoTe$_2$}},}\ }\href {https://doi.org/10.1103/PhysRevLett.126.106602} {\bibfield  {journal} {\bibinfo  {journal} {Physical Review Letters}\ }\textbf {\bibinfo {volume} {126}},\ \bibinfo {pages} {106602} (\bibinfo {year} {2021})}\BibitemShut {NoStop}%
\bibitem [{\citenamefont {Brumme}\ \emph {et~al.}(2015)\citenamefont {Brumme}, \citenamefont {Calandra},\ and\ \citenamefont {Mauri}}]{Brumme_PRB_2015}%
  \BibitemOpen
  \bibfield  {author} {\bibinfo {author} {\bibfnamefont {Thomas}\ \bibnamefont {Brumme}}, \bibinfo {author} {\bibfnamefont {Matteo}\ \bibnamefont {Calandra}}, \ and\ \bibinfo {author} {\bibfnamefont {Francesco}\ \bibnamefont {Mauri}},\ }\bibfield  {title} {\enquote {\bibinfo {title} {First-principles theory of field-effect doping in transition-metal dichalcogenides: Structural properties, electronic structure, hall coefficient, and electrical conductivity},}\ }\href@noop {} {\bibfield  {journal} {\bibinfo  {journal} {Physical Review B}\ }\textbf {\bibinfo {volume} {91}},\ \bibinfo {pages} {155436} (\bibinfo {year} {2015})}\BibitemShut {NoStop}%
\bibitem [{\citenamefont {Clark}\ \emph {et~al.}(2019)\citenamefont {Clark}, \citenamefont {Mazzola}, \citenamefont {Feng}, \citenamefont {Sunko}, \citenamefont {Markovi{\'c}}, \citenamefont {Bawden}, \citenamefont {Kim}, \citenamefont {King},\ and\ \citenamefont {Bahramy}}]{Clark_PRB2019}%
  \BibitemOpen
  \bibfield  {author} {\bibinfo {author} {\bibfnamefont {Oliver~Jon}\ \bibnamefont {Clark}}, \bibinfo {author} {\bibfnamefont {Federico}\ \bibnamefont {Mazzola}}, \bibinfo {author} {\bibfnamefont {Jiagui}\ \bibnamefont {Feng}}, \bibinfo {author} {\bibfnamefont {Veronika}\ \bibnamefont {Sunko}}, \bibinfo {author} {\bibfnamefont {Igor}\ \bibnamefont {Markovi{\'c}}}, \bibinfo {author} {\bibfnamefont {Lewis}\ \bibnamefont {Bawden}}, \bibinfo {author} {\bibfnamefont {Timur~K}\ \bibnamefont {Kim}}, \bibinfo {author} {\bibfnamefont {PDC}\ \bibnamefont {King}}, \ and\ \bibinfo {author} {\bibfnamefont {Mohammad~Saeed}\ \bibnamefont {Bahramy}},\ }\bibfield  {title} {\enquote {\bibinfo {title} {{Dual quantum confinement and anisotropic spin splitting in the multivalley semimetal PtSe$_{2}$}},}\ }\href {https://link.aps.org/doi/10.1103/PhysRevB.99.045438} {\bibfield  {journal} {\bibinfo  {journal} {Physical Review B}\ }\textbf {\bibinfo {volume} {99}},\ \bibinfo {pages} {045438} (\bibinfo {year} {2019})}\BibitemShut
  {NoStop}%
\bibitem [{\citenamefont {Boccuni}\ \emph {et~al.}(2024)\citenamefont {Boccuni}, \citenamefont {Peluzo}, \citenamefont {Bodo}, \citenamefont {Ambrogio}, \citenamefont {Maul}, \citenamefont {Mitoli}, \citenamefont {Vignale}, \citenamefont {Pittalis}, \citenamefont {Kraka}, \citenamefont {Desmarais} \emph {et~al.}}]{Boccuni_JPCL_2024}%
  \BibitemOpen
  \bibfield  {author} {\bibinfo {author} {\bibfnamefont {Alberto}\ \bibnamefont {Boccuni}}, \bibinfo {author} {\bibfnamefont {B{\'a}rbara Maria Teixeira~Costa}\ \bibnamefont {Peluzo}}, \bibinfo {author} {\bibfnamefont {Filippo}\ \bibnamefont {Bodo}}, \bibinfo {author} {\bibfnamefont {Giacomo}\ \bibnamefont {Ambrogio}}, \bibinfo {author} {\bibfnamefont {Jefferson}\ \bibnamefont {Maul}}, \bibinfo {author} {\bibfnamefont {Davide}\ \bibnamefont {Mitoli}}, \bibinfo {author} {\bibfnamefont {Giovanni}\ \bibnamefont {Vignale}}, \bibinfo {author} {\bibfnamefont {Stefano}\ \bibnamefont {Pittalis}}, \bibinfo {author} {\bibfnamefont {Elfi}\ \bibnamefont {Kraka}}, \bibinfo {author} {\bibfnamefont {Jacques~K}\ \bibnamefont {Desmarais}},  \emph {et~al.},\ }\bibfield  {title} {\enquote {\bibinfo {title} {Unveiling the role of spin currents on the giant rashba splitting in single-layer wse2},}\ }\href@noop {} {\bibfield  {journal} {\bibinfo  {journal} {The Journal of Physical Chemistry Letters}\ }\textbf {\bibinfo {volume}
  {15}},\ \bibinfo {pages} {7442--7448} (\bibinfo {year} {2024})}\BibitemShut {NoStop}%
\bibitem [{\citenamefont {Vom~Felde}\ \emph {et~al.}(1987)\citenamefont {Vom~Felde}, \citenamefont {Fink}, \citenamefont {B{\"u}che}, \citenamefont {Scheerer},\ and\ \citenamefont {N{\"u}cker}}]{vomFelde_EPL_1987}%
  \BibitemOpen
  \bibfield  {author} {\bibinfo {author} {\bibfnamefont {A}~\bibnamefont {Vom~Felde}}, \bibinfo {author} {\bibfnamefont {J}~\bibnamefont {Fink}}, \bibinfo {author} {\bibfnamefont {Th}~\bibnamefont {B{\"u}che}}, \bibinfo {author} {\bibfnamefont {B}~\bibnamefont {Scheerer}}, \ and\ \bibinfo {author} {\bibfnamefont {N}~\bibnamefont {N{\"u}cker}},\ }\bibfield  {title} {\enquote {\bibinfo {title} {Plasmons in the heavy alkali metals: strong deviations from {RPA}},}\ }\href {https://doi.org/10.1209/0295-5075/4/9/014} {\bibfield  {journal} {\bibinfo  {journal} {Europhysics Letters}\ }\textbf {\bibinfo {volume} {4}},\ \bibinfo {pages} {1037} (\bibinfo {year} {1987})}\BibitemShut {NoStop}%
\bibitem [{\citenamefont {Laturia}\ \emph {et~al.}(2018)\citenamefont {Laturia}, \citenamefont {Van~de Put},\ and\ \citenamefont {Vandenberghe}}]{Laturia_2DMatt_2018}%
  \BibitemOpen
  \bibfield  {author} {\bibinfo {author} {\bibfnamefont {Akash}\ \bibnamefont {Laturia}}, \bibinfo {author} {\bibfnamefont {Maarten~L}\ \bibnamefont {Van~de Put}}, \ and\ \bibinfo {author} {\bibfnamefont {William~G}\ \bibnamefont {Vandenberghe}},\ }\bibfield  {title} {\enquote {\bibinfo {title} {Dielectric properties of hexagonal boron nitride and transition metal dichalcogenides: from monolayer to bulk},}\ }\href {https://doi.org/10.1038/s41699-018-0050-x} {\bibfield  {journal} {\bibinfo  {journal} {npj 2D Materials and Applications}\ }\textbf {\bibinfo {volume} {2}},\ \bibinfo {pages} {6} (\bibinfo {year} {2018})}\BibitemShut {NoStop}%
\bibitem [{\citenamefont {Riley}\ \emph {et~al.}(2018)\citenamefont {Riley}, \citenamefont {Caruso}, \citenamefont {Verdi}, \citenamefont {Duffy}, \citenamefont {Watson}, \citenamefont {Bawden}, \citenamefont {Volckaert}, \citenamefont {van~der Laan}, \citenamefont {Hesjedal}, \citenamefont {Hoesch} \emph {et~al.}}]{Riley_NatCommun_2018}%
  \BibitemOpen
  \bibfield  {author} {\bibinfo {author} {\bibfnamefont {Jonathon~Mark}\ \bibnamefont {Riley}}, \bibinfo {author} {\bibfnamefont {F}~\bibnamefont {Caruso}}, \bibinfo {author} {\bibfnamefont {C}~\bibnamefont {Verdi}}, \bibinfo {author} {\bibfnamefont {LB}~\bibnamefont {Duffy}}, \bibinfo {author} {\bibfnamefont {Matthew~David}\ \bibnamefont {Watson}}, \bibinfo {author} {\bibfnamefont {Lewis}\ \bibnamefont {Bawden}}, \bibinfo {author} {\bibfnamefont {K}~\bibnamefont {Volckaert}}, \bibinfo {author} {\bibfnamefont {G}~\bibnamefont {van~der Laan}}, \bibinfo {author} {\bibfnamefont {T}~\bibnamefont {Hesjedal}}, \bibinfo {author} {\bibfnamefont {M}~\bibnamefont {Hoesch}},  \emph {et~al.},\ }\bibfield  {title} {\enquote {\bibinfo {title} {Crossover from lattice to plasmonic polarons of a spin-polarised electron gas in ferromagnetic {EuO}},}\ }\href {https://doi.org/10.1038/s41467-018-04749-w} {\bibfield  {journal} {\bibinfo  {journal} {Nature Communications}\ }\textbf {\bibinfo {volume} {9}},\ \bibinfo {pages} {2305}
  (\bibinfo {year} {2018})}\BibitemShut {NoStop}%
\bibitem [{\citenamefont {Ma}\ \emph {et~al.}(2020)\citenamefont {Ma}, \citenamefont {Cheng}, \citenamefont {Tian}, \citenamefont {Liu}, \citenamefont {Cui}, \citenamefont {Huang}, \citenamefont {Tan}, \citenamefont {Yang},\ and\ \citenamefont {Wang}}]{Ma_NanoLett_2020}%
  \BibitemOpen
  \bibfield  {author} {\bibinfo {author} {\bibfnamefont {Xiaochuan}\ \bibnamefont {Ma}}, \bibinfo {author} {\bibfnamefont {Zhengwang}\ \bibnamefont {Cheng}}, \bibinfo {author} {\bibfnamefont {Mingyang}\ \bibnamefont {Tian}}, \bibinfo {author} {\bibfnamefont {Xiaofeng}\ \bibnamefont {Liu}}, \bibinfo {author} {\bibfnamefont {Xuefeng}\ \bibnamefont {Cui}}, \bibinfo {author} {\bibfnamefont {Yaobo}\ \bibnamefont {Huang}}, \bibinfo {author} {\bibfnamefont {Shijing}\ \bibnamefont {Tan}}, \bibinfo {author} {\bibfnamefont {Jinlong}\ \bibnamefont {Yang}}, \ and\ \bibinfo {author} {\bibfnamefont {Bing}\ \bibnamefont {Wang}},\ }\bibfield  {title} {\enquote {\bibinfo {title} {Formation of plasmonic polarons in highly electron-doped anatase {TiO$_{2}$}},}\ }\href {https://doi.org/10.1021/acs.nanolett.0c03802} {\bibfield  {journal} {\bibinfo  {journal} {Nano Letters}\ }\textbf {\bibinfo {volume} {21}},\ \bibinfo {pages} {430--436} (\bibinfo {year} {2020})}\BibitemShut {NoStop}%
\bibitem [{\citenamefont {Caruso}\ \emph {et~al.}(2021)\citenamefont {Caruso}, \citenamefont {Amsalem}, \citenamefont {Ma}, \citenamefont {Aljarb}, \citenamefont {Schultz}, \citenamefont {Zacharias}, \citenamefont {Tung}, \citenamefont {Koch},\ and\ \citenamefont {Draxl}}]{Caruso_PRB2021}%
  \BibitemOpen
  \bibfield  {author} {\bibinfo {author} {\bibfnamefont {Fabio}\ \bibnamefont {Caruso}}, \bibinfo {author} {\bibfnamefont {Patrick}\ \bibnamefont {Amsalem}}, \bibinfo {author} {\bibfnamefont {Jie}\ \bibnamefont {Ma}}, \bibinfo {author} {\bibfnamefont {Areej}\ \bibnamefont {Aljarb}}, \bibinfo {author} {\bibfnamefont {Thorsten}\ \bibnamefont {Schultz}}, \bibinfo {author} {\bibfnamefont {Marios}\ \bibnamefont {Zacharias}}, \bibinfo {author} {\bibfnamefont {Vincent}\ \bibnamefont {Tung}}, \bibinfo {author} {\bibfnamefont {Norbert}\ \bibnamefont {Koch}}, \ and\ \bibinfo {author} {\bibfnamefont {Claudia}\ \bibnamefont {Draxl}},\ }\bibfield  {title} {\enquote {\bibinfo {title} {Two-dimensional plasmonic polarons in $n$-doped monolayer {MoS$_{2}$}},}\ }\href {\doibase 10.1103/PhysRevB.103.205152} {\bibfield  {journal} {\bibinfo  {journal} {Physical Review B}\ }\textbf {\bibinfo {volume} {103}},\ \bibinfo {pages} {205152} (\bibinfo {year} {2021})}\BibitemShut {NoStop}%
\bibitem [{\citenamefont {Xiang}\ \emph {et~al.}(2023)\citenamefont {Xiang}, \citenamefont {Ma}, \citenamefont {Gao}, \citenamefont {Guo}, \citenamefont {Huang}, \citenamefont {Xing}, \citenamefont {Tan}, \citenamefont {Zhao}, \citenamefont {Wang},\ and\ \citenamefont {Shao}}]{Xiang_JPCL_2023}%
  \BibitemOpen
  \bibfield  {author} {\bibinfo {author} {\bibfnamefont {Miaomiao}\ \bibnamefont {Xiang}}, \bibinfo {author} {\bibfnamefont {Xiaochuan}\ \bibnamefont {Ma}}, \bibinfo {author} {\bibfnamefont {Chang}\ \bibnamefont {Gao}}, \bibinfo {author} {\bibfnamefont {Ziyang}\ \bibnamefont {Guo}}, \bibinfo {author} {\bibfnamefont {Chenxi}\ \bibnamefont {Huang}}, \bibinfo {author} {\bibfnamefont {Yue}\ \bibnamefont {Xing}}, \bibinfo {author} {\bibfnamefont {Shijing}\ \bibnamefont {Tan}}, \bibinfo {author} {\bibfnamefont {Jin}\ \bibnamefont {Zhao}}, \bibinfo {author} {\bibfnamefont {Bing}\ \bibnamefont {Wang}}, \ and\ \bibinfo {author} {\bibfnamefont {Xiang}\ \bibnamefont {Shao}},\ }\bibfield  {title} {\enquote {\bibinfo {title} {Revealing the polaron state at the {MoS$_2$}/{TiO$_2$} interface},}\ }\href {https://doi.org/10.1021/acs.jpclett.2c03856} {\bibfield  {journal} {\bibinfo  {journal} {The Journal of Physical Chemistry Letters}\ }\textbf {\bibinfo {volume} {14}},\ \bibinfo {pages} {3360--3367} (\bibinfo {year}
  {2023})}\BibitemShut {NoStop}%
\bibitem [{\citenamefont {Emeis}\ \emph {et~al.}(2023)\citenamefont {Emeis}, \citenamefont {Mahatha}, \citenamefont {Rohlf}, \citenamefont {Rossnagel},\ and\ \citenamefont {Caruso}}]{Emeis_PRB2023}%
  \BibitemOpen
  \bibfield  {author} {\bibinfo {author} {\bibfnamefont {Christoph}\ \bibnamefont {Emeis}}, \bibinfo {author} {\bibfnamefont {Sanjoy~Kr}\ \bibnamefont {Mahatha}}, \bibinfo {author} {\bibfnamefont {Sebastian}\ \bibnamefont {Rohlf}}, \bibinfo {author} {\bibfnamefont {Kai}\ \bibnamefont {Rossnagel}}, \ and\ \bibinfo {author} {\bibfnamefont {Fabio}\ \bibnamefont {Caruso}},\ }\bibfield  {title} {\enquote {\bibinfo {title} {Plasmonic polarons induced by alkali-atom deposition in hafnium disulfide {1T-HfS$_{2}$}},}\ }\href {\doibase 10.1103/PhysRevB.108.155149} {\bibfield  {journal} {\bibinfo  {journal} {Physical Review B}\ }\textbf {\bibinfo {volume} {108}},\ \bibinfo {pages} {155149} (\bibinfo {year} {2023})}\BibitemShut {NoStop}%
\bibitem [{\citenamefont {Ulstrup}\ \emph {et~al.}(2023)\citenamefont {Ulstrup}, \citenamefont {Miwa}, \citenamefont {Jones}, \citenamefont {McCreary}, \citenamefont {Robinson}, \citenamefont {Jonker}, \citenamefont {Singh}, \citenamefont {Koch}, \citenamefont {Rotenberg}, \citenamefont {Bostwick} \emph {et~al.}}]{Ulstrup_arxiv_2308_16509}%
  \BibitemOpen
  \bibfield  {author} {\bibinfo {author} {\bibfnamefont {S{\o}ren}\ \bibnamefont {Ulstrup}}, \bibinfo {author} {\bibfnamefont {Jill~A}\ \bibnamefont {Miwa}}, \bibinfo {author} {\bibfnamefont {Alfred~JH}\ \bibnamefont {Jones}}, \bibinfo {author} {\bibfnamefont {Kathleen~M}\ \bibnamefont {McCreary}}, \bibinfo {author} {\bibfnamefont {Jeremy~T}\ \bibnamefont {Robinson}}, \bibinfo {author} {\bibfnamefont {Berend~T}\ \bibnamefont {Jonker}}, \bibinfo {author} {\bibfnamefont {Simranjeet}\ \bibnamefont {Singh}}, \bibinfo {author} {\bibfnamefont {Roland~J}\ \bibnamefont {Koch}}, \bibinfo {author} {\bibfnamefont {Eli}\ \bibnamefont {Rotenberg}}, \bibinfo {author} {\bibfnamefont {Aaron}\ \bibnamefont {Bostwick}},  \emph {et~al.},\ }\bibfield  {title} {\enquote {\bibinfo {title} {{Discovery of interlayer plasmon polaron in graphene/WS$_2$ heterostructures}},}\ }\href {https://arxiv.org/abs/2308.16509} {\bibfield  {journal} {\bibinfo  {journal} {arXiv preprint arXiv:2308.16509}\ } (\bibinfo {year} {2023})}\BibitemShut
  {NoStop}%
\bibitem [{\citenamefont {Shi}\ \emph {et~al.}(2015)\citenamefont {Shi}, \citenamefont {Ye}, \citenamefont {Zhang}, \citenamefont {Suzuki}, \citenamefont {Yoshida}, \citenamefont {Miyazaki}, \citenamefont {Inoue}, \citenamefont {Saito},\ and\ \citenamefont {Iwasa}}]{Shi_SciRep_2015}%
  \BibitemOpen
  \bibfield  {author} {\bibinfo {author} {\bibfnamefont {Wu}~\bibnamefont {Shi}}, \bibinfo {author} {\bibfnamefont {Jianting}\ \bibnamefont {Ye}}, \bibinfo {author} {\bibfnamefont {Yijin}\ \bibnamefont {Zhang}}, \bibinfo {author} {\bibfnamefont {Ryuji}\ \bibnamefont {Suzuki}}, \bibinfo {author} {\bibfnamefont {Masaro}\ \bibnamefont {Yoshida}}, \bibinfo {author} {\bibfnamefont {Jun}\ \bibnamefont {Miyazaki}}, \bibinfo {author} {\bibfnamefont {Naoko}\ \bibnamefont {Inoue}}, \bibinfo {author} {\bibfnamefont {Yu}~\bibnamefont {Saito}}, \ and\ \bibinfo {author} {\bibfnamefont {Yoshihiro}\ \bibnamefont {Iwasa}},\ }\bibfield  {title} {\enquote {\bibinfo {title} {Superconductivity series in transition metal dichalcogenides by ionic gating},}\ }\href {https://doi.org/10.1038/srep12534} {\bibfield  {journal} {\bibinfo  {journal} {Scientific Reports}\ }\textbf {\bibinfo {volume} {5}},\ \bibinfo {pages} {12534} (\bibinfo {year} {2015})}\BibitemShut {NoStop}%
\bibitem [{\citenamefont {Taniguchi}\ \emph {et~al.}(2012)\citenamefont {Taniguchi}, \citenamefont {Matsumoto}, \citenamefont {Shimotani},\ and\ \citenamefont {Takagi}}]{Taniguchi_APL_2012}%
  \BibitemOpen
  \bibfield  {author} {\bibinfo {author} {\bibfnamefont {Kouji}\ \bibnamefont {Taniguchi}}, \bibinfo {author} {\bibfnamefont {Akiyo}\ \bibnamefont {Matsumoto}}, \bibinfo {author} {\bibfnamefont {Hidekazu}\ \bibnamefont {Shimotani}}, \ and\ \bibinfo {author} {\bibfnamefont {Hidenori}\ \bibnamefont {Takagi}},\ }\bibfield  {title} {\enquote {\bibinfo {title} {Electric-field-induced superconductivity at 9.4 k in a layered transition metal disulphide {MoS$_{2}$}},}\ }\href {http://dx.doi.org/10.1063/1.4740268} {\bibfield  {journal} {\bibinfo  {journal} {Applied Physics Letters}\ }\textbf {\bibinfo {volume} {101}} (\bibinfo {year} {2012})}\BibitemShut {NoStop}%
\bibitem [{\citenamefont {Wu}\ \emph {et~al.}(2021)\citenamefont {Wu}, \citenamefont {Lin}, \citenamefont {Xiong}, \citenamefont {Li}, \citenamefont {Luo}, \citenamefont {Chen},\ and\ \citenamefont {Zheng}}]{Wu_PRB_2021}%
  \BibitemOpen
  \bibfield  {author} {\bibinfo {author} {\bibfnamefont {Danhong}\ \bibnamefont {Wu}}, \bibinfo {author} {\bibfnamefont {Yiping}\ \bibnamefont {Lin}}, \bibinfo {author} {\bibfnamefont {Lingxiao}\ \bibnamefont {Xiong}}, \bibinfo {author} {\bibfnamefont {Junjie}\ \bibnamefont {Li}}, \bibinfo {author} {\bibfnamefont {Tiantian}\ \bibnamefont {Luo}}, \bibinfo {author} {\bibfnamefont {Deyi}\ \bibnamefont {Chen}}, \ and\ \bibinfo {author} {\bibfnamefont {Feipeng}\ \bibnamefont {Zheng}},\ }\bibfield  {title} {\enquote {\bibinfo {title} {Enhanced superconductivity in bilayer {PtTe$_{2}$} by alkali-metal intercalations},}\ }\href {https://doi.org/10.1103/PhysRevB.103.224502} {\bibfield  {journal} {\bibinfo  {journal} {Physical Review B}\ }\textbf {\bibinfo {volume} {103}},\ \bibinfo {pages} {224502} (\bibinfo {year} {2021})}\BibitemShut {NoStop}%
\bibitem [{\citenamefont {Saito}\ \emph {et~al.}(2016)\citenamefont {Saito}, \citenamefont {Nakamura}, \citenamefont {Bahramy}, \citenamefont {Kohama}, \citenamefont {Ye}, \citenamefont {Kasahara}, \citenamefont {Nakagawa}, \citenamefont {Onga}, \citenamefont {Tokunaga}, \citenamefont {Nojima} \emph {et~al.}}]{Saito_NaturePhys_2016}%
  \BibitemOpen
  \bibfield  {author} {\bibinfo {author} {\bibfnamefont {Yu}~\bibnamefont {Saito}}, \bibinfo {author} {\bibfnamefont {Yasuharu}\ \bibnamefont {Nakamura}}, \bibinfo {author} {\bibfnamefont {Mohammad~Saeed}\ \bibnamefont {Bahramy}}, \bibinfo {author} {\bibfnamefont {Yoshimitsu}\ \bibnamefont {Kohama}}, \bibinfo {author} {\bibfnamefont {Jianting}\ \bibnamefont {Ye}}, \bibinfo {author} {\bibfnamefont {Yuichi}\ \bibnamefont {Kasahara}}, \bibinfo {author} {\bibfnamefont {Yuji}\ \bibnamefont {Nakagawa}}, \bibinfo {author} {\bibfnamefont {Masaru}\ \bibnamefont {Onga}}, \bibinfo {author} {\bibfnamefont {Masashi}\ \bibnamefont {Tokunaga}}, \bibinfo {author} {\bibfnamefont {Tsutomu}\ \bibnamefont {Nojima}},  \emph {et~al.},\ }\bibfield  {title} {\enquote {\bibinfo {title} {Superconductivity protected by spin--valley locking in ion-gated {MoS$_{2}$}},}\ }\href {https://doi.org/10.1038/nphys3580} {\bibfield  {journal} {\bibinfo  {journal} {Nature Physics}\ }\textbf {\bibinfo {volume} {12}},\ \bibinfo {pages} {144--149}
  (\bibinfo {year} {2016})}\BibitemShut {NoStop}%
\bibitem [{\citenamefont {Zhou}\ \emph {et~al.}(2016)\citenamefont {Zhou}, \citenamefont {Yuan}, \citenamefont {Jiang},\ and\ \citenamefont {Law}}]{Zhou_PRB_2016}%
  \BibitemOpen
  \bibfield  {author} {\bibinfo {author} {\bibfnamefont {Benjamin~T.}\ \bibnamefont {Zhou}}, \bibinfo {author} {\bibfnamefont {Noah F.~Q.}\ \bibnamefont {Yuan}}, \bibinfo {author} {\bibfnamefont {Hong-Liang}\ \bibnamefont {Jiang}}, \ and\ \bibinfo {author} {\bibfnamefont {K.~T.}\ \bibnamefont {Law}},\ }\bibfield  {title} {\enquote {\bibinfo {title} {Ising superconductivity and {M}ajorana fermions in transition-metal dichalcogenides},}\ }\href {\doibase 10.1103/PhysRevB.93.180501} {\bibfield  {journal} {\bibinfo  {journal} {Phys. Rev. B}\ }\textbf {\bibinfo {volume} {93}},\ \bibinfo {pages} {180501(R)} (\bibinfo {year} {2016})}\BibitemShut {NoStop}%
\bibitem [{\citenamefont {Ili\ifmmode~\acute{c}\else \'{c}\fi{}}\ \emph {et~al.}(2017)\citenamefont {Ili\ifmmode~\acute{c}\else \'{c}\fi{}}, \citenamefont {Meyer},\ and\ \citenamefont {Houzet}}]{Ilic_PRL_2017}%
  \BibitemOpen
  \bibfield  {author} {\bibinfo {author} {\bibfnamefont {Stefan}\ \bibnamefont {Ili\ifmmode~\acute{c}\else \'{c}\fi{}}}, \bibinfo {author} {\bibfnamefont {Julia~S.}\ \bibnamefont {Meyer}}, \ and\ \bibinfo {author} {\bibfnamefont {Manuel}\ \bibnamefont {Houzet}},\ }\bibfield  {title} {\enquote {\bibinfo {title} {Enhancement of the upper critical field in disordered transition metal dichalcogenide monolayers},}\ }\href {\doibase 10.1103/PhysRevLett.119.117001} {\bibfield  {journal} {\bibinfo  {journal} {Phys. Rev. Lett.}\ }\textbf {\bibinfo {volume} {119}},\ \bibinfo {pages} {117001} (\bibinfo {year} {2017})}\BibitemShut {NoStop}%
\bibitem [{\citenamefont {Wang}\ \emph {et~al.}(2019)\citenamefont {Wang}, \citenamefont {Lian}, \citenamefont {Guo}, \citenamefont {Mao}, \citenamefont {Zhang}, \citenamefont {Zhang}, \citenamefont {Gu}, \citenamefont {Xu},\ and\ \citenamefont {Duan}}]{Wang_PRL_2019}%
  \BibitemOpen
  \bibfield  {author} {\bibinfo {author} {\bibfnamefont {Chong}\ \bibnamefont {Wang}}, \bibinfo {author} {\bibfnamefont {Biao}\ \bibnamefont {Lian}}, \bibinfo {author} {\bibfnamefont {Xiaomi}\ \bibnamefont {Guo}}, \bibinfo {author} {\bibfnamefont {Jiahao}\ \bibnamefont {Mao}}, \bibinfo {author} {\bibfnamefont {Zetao}\ \bibnamefont {Zhang}}, \bibinfo {author} {\bibfnamefont {Ding}\ \bibnamefont {Zhang}}, \bibinfo {author} {\bibfnamefont {Bing-Lin}\ \bibnamefont {Gu}}, \bibinfo {author} {\bibfnamefont {Yong}\ \bibnamefont {Xu}}, \ and\ \bibinfo {author} {\bibfnamefont {Wenhui}\ \bibnamefont {Duan}},\ }\bibfield  {title} {\enquote {\bibinfo {title} {Type-{II} {I}sing superconductivity in two-dimensional materials with spin-orbit coupling},}\ }\href {\doibase 10.1103/PhysRevLett.123.126402} {\bibfield  {journal} {\bibinfo  {journal} {Physical review letters}\ }\textbf {\bibinfo {volume} {123}},\ \bibinfo {pages} {126402} (\bibinfo {year} {2019})}\BibitemShut {NoStop}%
\bibitem [{\citenamefont {Liu}\ \emph {et~al.}(2020)\citenamefont {Liu}, \citenamefont {Liu}, \citenamefont {Zhang},\ and\ \citenamefont {Xie}}]{Liu_PRB_2020}%
  \BibitemOpen
  \bibfield  {author} {\bibinfo {author} {\bibfnamefont {Hongchao}\ \bibnamefont {Liu}}, \bibinfo {author} {\bibfnamefont {Haiwen}\ \bibnamefont {Liu}}, \bibinfo {author} {\bibfnamefont {Ding}\ \bibnamefont {Zhang}}, \ and\ \bibinfo {author} {\bibfnamefont {X.~C.}\ \bibnamefont {Xie}},\ }\bibfield  {title} {\enquote {\bibinfo {title} {Microscopic theory of in-plane critical field in two-dimensional {I}sing superconducting systems},}\ }\href {\doibase 10.1103/PhysRevB.102.174510} {\bibfield  {journal} {\bibinfo  {journal} {Phys. Rev. B}\ }\textbf {\bibinfo {volume} {102}},\ \bibinfo {pages} {174510} (\bibinfo {year} {2020})}\BibitemShut {NoStop}%
\bibitem [{\citenamefont {Wickramaratne}\ \emph {et~al.}(2020)\citenamefont {Wickramaratne}, \citenamefont {Khmelevskyi}, \citenamefont {Agterberg},\ and\ \citenamefont {Mazin}}]{Wickramaratne_PRX_2020}%
  \BibitemOpen
  \bibfield  {author} {\bibinfo {author} {\bibfnamefont {Darshana}\ \bibnamefont {Wickramaratne}}, \bibinfo {author} {\bibfnamefont {Sergii}\ \bibnamefont {Khmelevskyi}}, \bibinfo {author} {\bibfnamefont {Daniel~F.}\ \bibnamefont {Agterberg}}, \ and\ \bibinfo {author} {\bibfnamefont {I.~I.}\ \bibnamefont {Mazin}},\ }\bibfield  {title} {\enquote {\bibinfo {title} {Ising superconductivity and magnetism in ${\mathrm{nbse}}_{2}$},}\ }\href {\doibase 10.1103/PhysRevX.10.041003} {\bibfield  {journal} {\bibinfo  {journal} {Phys. Rev. X}\ }\textbf {\bibinfo {volume} {10}},\ \bibinfo {pages} {041003} (\bibinfo {year} {2020})}\BibitemShut {NoStop}%
\bibitem [{\citenamefont {Semenov}(2024)}]{Semenov_JETP_Lrtt_2024}%
  \BibitemOpen
  \bibfield  {author} {\bibinfo {author} {\bibfnamefont {A.~G.}\ \bibnamefont {Semenov}},\ }\bibfield  {title} {\enquote {\bibinfo {title} {Pairing and collective excitations in ising superconductors},}\ }\href {\doibase 10.1134/S0021364023603810} {\bibfield  {journal} {\bibinfo  {journal} {JETP Letters}\ }\textbf {\bibinfo {volume} {119}},\ \bibinfo {pages} {46--52} (\bibinfo {year} {2024})}\BibitemShut {NoStop}%
\bibitem [{\citenamefont {Zhao}\ \emph {et~al.}(2020)\citenamefont {Zhao}, \citenamefont {Yu}, \citenamefont {Zhong}, \citenamefont {R{\"o}sner}, \citenamefont {Katsnelson},\ and\ \citenamefont {Yuan}}]{Zhao_NJP_2020}%
  \BibitemOpen
  \bibfield  {author} {\bibinfo {author} {\bibfnamefont {Peiliang}\ \bibnamefont {Zhao}}, \bibinfo {author} {\bibfnamefont {Jin}\ \bibnamefont {Yu}}, \bibinfo {author} {\bibfnamefont {H}~\bibnamefont {Zhong}}, \bibinfo {author} {\bibfnamefont {Malte}\ \bibnamefont {R{\"o}sner}}, \bibinfo {author} {\bibfnamefont {Mikhail~I}\ \bibnamefont {Katsnelson}}, \ and\ \bibinfo {author} {\bibfnamefont {Shengjun}\ \bibnamefont {Yuan}},\ }\bibfield  {title} {\enquote {\bibinfo {title} {Electronic and optical properties of transition metal dichalcogenides under symmetric and asymmetric field-effect doping},}\ }\href {\doibase 10.1088/1367-2630/aba8d2} {\bibfield  {journal} {\bibinfo  {journal} {New Journal of Physics}\ }\textbf {\bibinfo {volume} {22}},\ \bibinfo {pages} {083072} (\bibinfo {year} {2020})}\BibitemShut {NoStop}%
\bibitem [{\citenamefont {Covaci}\ and\ \citenamefont {Berciu}(2009)}]{Covaci_PRL_2009}%
  \BibitemOpen
  \bibfield  {author} {\bibinfo {author} {\bibfnamefont {Lucian}\ \bibnamefont {Covaci}}\ and\ \bibinfo {author} {\bibfnamefont {Mona}\ \bibnamefont {Berciu}},\ }\bibfield  {title} {\enquote {\bibinfo {title} {Polaron formation in the presence of rashba spin-orbit coupling: implications for spintronics},}\ }\href {https://doi.org/10.1103/PhysRevLett.102.186403} {\bibfield  {journal} {\bibinfo  {journal} {Physical Review Letters}\ }\textbf {\bibinfo {volume} {102}},\ \bibinfo {pages} {186403} (\bibinfo {year} {2009})}\BibitemShut {NoStop}%
\bibitem [{\citenamefont {Legma}\ \emph {et~al.}(1993)\citenamefont {Legma}, \citenamefont {Vacquier},\ and\ \citenamefont {Casalot}}]{Legma_JCG1993}%
  \BibitemOpen
  \bibfield  {author} {\bibinfo {author} {\bibfnamefont {J.B.}\ \bibnamefont {Legma}}, \bibinfo {author} {\bibfnamefont {G.}~\bibnamefont {Vacquier}}, \ and\ \bibinfo {author} {\bibfnamefont {A.}~\bibnamefont {Casalot}},\ }\bibfield  {title} {\enquote {\bibinfo {title} {Chemical vapour transport of molybdenum and tungsten diselenides by various transport agents},}\ }\href {\doibase https://doi.org/10.1016/0022-0248(93)90859-U} {\bibfield  {journal} {\bibinfo  {journal} {Journal of Crystal Growth}\ }\textbf {\bibinfo {volume} {130}},\ \bibinfo {pages} {253--258} (\bibinfo {year} {1993})}\BibitemShut {NoStop}%
\bibitem [{\citenamefont {Hoesch}\ \emph {et~al.}(2017)\citenamefont {Hoesch}, \citenamefont {Kim}, \citenamefont {Dudin}, \citenamefont {Wang}, \citenamefont {Scott}, \citenamefont {Harris}, \citenamefont {Patel}, \citenamefont {Matthews}, \citenamefont {Hawkins}, \citenamefont {Alcock} \emph {et~al.}}]{hoesch2017facility}%
  \BibitemOpen
  \bibfield  {author} {\bibinfo {author} {\bibfnamefont {M}~\bibnamefont {Hoesch}}, \bibinfo {author} {\bibfnamefont {TK}~\bibnamefont {Kim}}, \bibinfo {author} {\bibfnamefont {P}~\bibnamefont {Dudin}}, \bibinfo {author} {\bibfnamefont {H}~\bibnamefont {Wang}}, \bibinfo {author} {\bibfnamefont {S}~\bibnamefont {Scott}}, \bibinfo {author} {\bibfnamefont {P}~\bibnamefont {Harris}}, \bibinfo {author} {\bibfnamefont {S}~\bibnamefont {Patel}}, \bibinfo {author} {\bibfnamefont {M}~\bibnamefont {Matthews}}, \bibinfo {author} {\bibfnamefont {D}~\bibnamefont {Hawkins}}, \bibinfo {author} {\bibfnamefont {SG}~\bibnamefont {Alcock}},  \emph {et~al.},\ }\bibfield  {title} {\enquote {\bibinfo {title} {A facility for the analysis of the electronic structures of solids and their surfaces by synchrotron radiation photoelectron spectroscopy},}\ }\href {\doibase https://doi.org/10.1063/1.4973562} {\bibfield  {journal} {\bibinfo  {journal} {Review of Scientific Instruments}\ }\textbf {\bibinfo {volume} {88}},\ \bibinfo {pages}
  {013106} (\bibinfo {year} {2017})}\BibitemShut {NoStop}%
\bibitem [{\citenamefont {Blaha}\ \emph {et~al.}(2020)\citenamefont {Blaha}, \citenamefont {Schwarz}, \citenamefont {Tran}, \citenamefont {Laskowski}, \citenamefont {Madsen},\ and\ \citenamefont {Marks}}]{Blaha_JCP_2020wien2k}%
  \BibitemOpen
  \bibfield  {author} {\bibinfo {author} {\bibfnamefont {Peter}\ \bibnamefont {Blaha}}, \bibinfo {author} {\bibfnamefont {Karlheinz}\ \bibnamefont {Schwarz}}, \bibinfo {author} {\bibfnamefont {Fabien}\ \bibnamefont {Tran}}, \bibinfo {author} {\bibfnamefont {Robert}\ \bibnamefont {Laskowski}}, \bibinfo {author} {\bibfnamefont {Georg~KH}\ \bibnamefont {Madsen}}, \ and\ \bibinfo {author} {\bibfnamefont {Laurence~D}\ \bibnamefont {Marks}},\ }\bibfield  {title} {\enquote {\bibinfo {title} {{WIEN2k: An APW+ lo program for calculating the properties of solids}},}\ }\href {https://doi.org/10.1063/1.5143061} {\bibfield  {journal} {\bibinfo  {journal} {The Journal of Chemical Physics}\ }\textbf {\bibinfo {volume} {152}},\ \bibinfo {pages} {074101} (\bibinfo {year} {2020})}\BibitemShut {NoStop}%
\bibitem [{\citenamefont {El~Youbi}\ \emph {et~al.}(2020)\citenamefont {El~Youbi}, \citenamefont {Jung}, \citenamefont {Mukherjee}, \citenamefont {Fanciulli}, \citenamefont {Schusser}, \citenamefont {Heckmann}, \citenamefont {Richter}, \citenamefont {Min{\'a}r}, \citenamefont {Hricovini}, \citenamefont {Watson},\ and\ \citenamefont {Cacho}}]{Zak_PRB2020}%
  \BibitemOpen
  \bibfield  {author} {\bibinfo {author} {\bibfnamefont {Zakariae}\ \bibnamefont {El~Youbi}}, \bibinfo {author} {\bibfnamefont {Sung~Won}\ \bibnamefont {Jung}}, \bibinfo {author} {\bibfnamefont {Saumya}\ \bibnamefont {Mukherjee}}, \bibinfo {author} {\bibfnamefont {Mauro}\ \bibnamefont {Fanciulli}}, \bibinfo {author} {\bibfnamefont {Jakub}\ \bibnamefont {Schusser}}, \bibinfo {author} {\bibfnamefont {Olivier}\ \bibnamefont {Heckmann}}, \bibinfo {author} {\bibfnamefont {Christine}\ \bibnamefont {Richter}}, \bibinfo {author} {\bibfnamefont {J{\'a}n}\ \bibnamefont {Min{\'a}r}}, \bibinfo {author} {\bibfnamefont {Karol}\ \bibnamefont {Hricovini}}, \bibinfo {author} {\bibfnamefont {Matthew~D.}\ \bibnamefont {Watson}}, \ and\ \bibinfo {author} {\bibfnamefont {Cephise}\ \bibnamefont {Cacho}},\ }\bibfield  {title} {\enquote {\bibinfo {title} {{Bulk and surface electronic states in the dosed semimetallic HfTe$_2$}},}\ }\href {\doibase 10.1103/PhysRevB.101.235431} {\bibfield  {journal} {\bibinfo  {journal} {Physical
  Review B}\ }\textbf {\bibinfo {volume} {101}},\ \bibinfo {pages} {235431} (\bibinfo {year} {2020})}\BibitemShut {NoStop}%
\end{thebibliography}%

\clearpage
\begin{widetext} 
\section*{Supplementary} \label{sec:supplementary}

\counterwithin{figure}{section}
\renewcommand{\thefigure}{S\arabic{figure}}
\setcounter{figure}{0}
\subsection*{Density Functional Theory (DFT) calculations}

Band structure DFT calculations including spin-orbit coupling show that bottom of the conduction band at $\Sigma$-valley is lower for bulk MoSe$_2$ then for bulk MoS$_2$ and lower then bottom of the conduction band at ${\rm K}$-valley. 
This makes MoSe$_2$ a much better choice to study polaron formation at $\Sigma$-valley.
Importantly, in case of broken inversion symmetry, due to electric field built up on the surface caused by the alkali metal dosing, one expect corresponding spin-splitting at $\Sigma$-valley, like in the case of monolayer MoSe$_2$.

\begin{figure}[hbt]
\centering
\includegraphics[width=\linewidth]{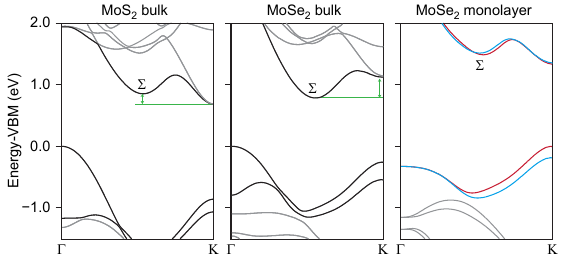}
\caption{\textbf{DFT+SOC calculated electronic structure of bulk MoS$_2$, bulk MoSe$_2$ and monolayer MoSe$_2$.}}
\label{Fig_S_DFT}
\end{figure}

\clearpage
\subsection*{The evolution of MoSe$_2$ electronic structure with Rb dosing.}
Photoemission data for conduction (Fig.~\ref{Fig_S_CB}) and valence (Fig.~\ref{Fig_S_VB}) bands together with corresponding Se 3d core levels data (Fig.~\ref{Fig_S_XPS}) were measured for all Rb dosing sequences.

\begin{figure}[hbt]
\centering
\includegraphics[width=\linewidth]{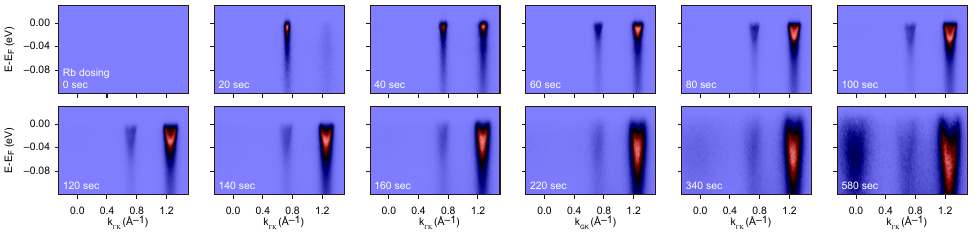}
\caption{\textbf{Conduction band electronic structure at the Fermi-level of Rb dosed MoSe$_2$ surface.}}
\label{Fig_S_CB}
\end{figure}

\begin{figure}[h]
\centering
\includegraphics[width=\linewidth]{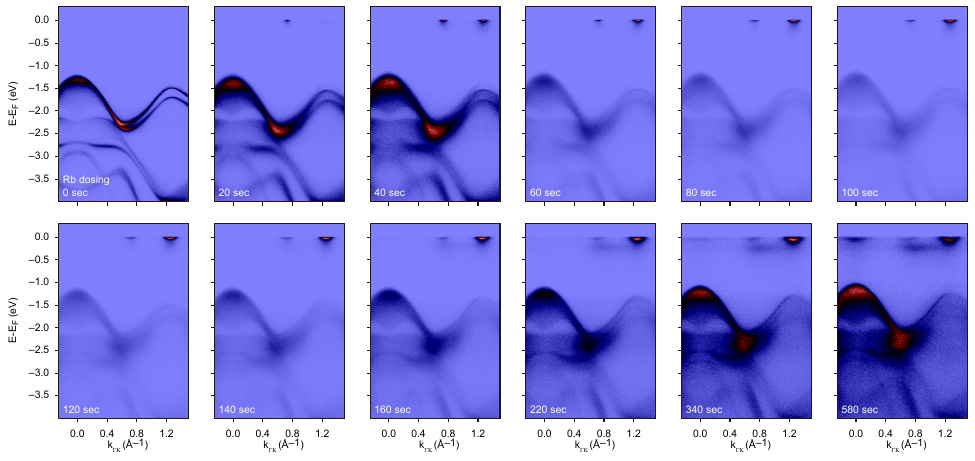}
\caption{\textbf{Valence band electronic structure of Rb dosed MoSe$_2$ surface.}}
\label{Fig_S_VB}
\end{figure}

\begin{figure}[h]
\centering
\includegraphics[width=\linewidth]{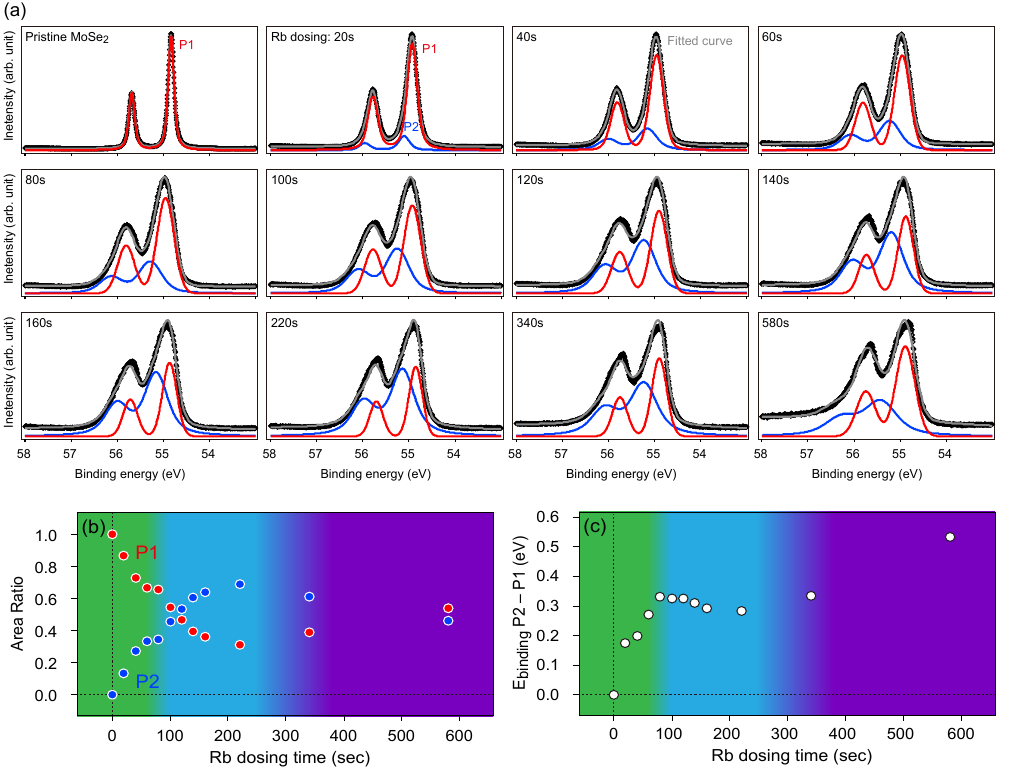}
\caption{\textbf{Se 3p core-levels with Rb dosing.}
(a) Se 3d$_{5/2}$/3d$_{3/2}$ XPS data with Voight profile doublets fit for various dosing times. Red and blue curves indicate bulk (P1) and surface (P2) components; (b) Ratio of the bulk and surface components versus Rb dosing time; (c) Energy difference between bulk and surface components versus Rb dosing time; The background colour in (b) and (c) indicates surface dosing, intercalation and Rb metallisation regimes.}
\label{Fig_S_XPS}
\end{figure}

A single 3d$_{5/2}$/3d$_{3/2}$ doublet for a pristine MoSe$_2$ crystal splits into two components with Rb dosing as shown in Fig.~\ref{Fig_S_XPS}(a).
Similar to the case of HfTe$_2$~\cite{Zak_PRB2020} one could fit these core level data assuming bulk (P1) and surface (P2) components due to the electron doping of the MoSe$_2$ surface.
Obtained peak area ratio and energy differences (or effective chemical shift) are shown in Fig.~\ref{Fig_S_XPS}(b,c).
Below critical dosing of 120\,s one observes increase of the difference between bulk and surface doublets binding energies with dosing time.
This corresponds to the gradual accumulation of Rb atoms, increased electron doping of the MoSe$_2$ surface, and corresponding increase of effective electric field. Beyond this critical dosing time, Rb atoms penetrate the van der Waals gap underneath the top layer, thereby physically and electronically isolating the topmost MoSe$_2$ layer. This clearly marks the second crossover from surface doping to the interlayer intercalation regime.

\clearpage
\subsection*{Doping evolution of the band gap in MoSe$_2$}

Similar to the previously published ARPES data of alkali-dosed MoSe$_2$~\cite{Kim_SciRep2017, Kang_NanoLett2017} we observe superposition of the bulk and surface electronic states.
This is especially evident at top of valence band at ${\rm K}$-valley where one detects weak surface related features at lower binding energies than bulk ones.
At higher doping levels this might lead to a shift of the valence band maxima (VBM) from $\Gamma$ to ${\rm K}$ valley and corresponding transition from indirect to direct band gap electronic structure.
This has been clearly observed by ARPES in the case of MoTe$_2$~\cite{Kang_NanoLett2017}.
However, in the case of MoSe$_2$ this is less obvious from the ARPES data~\cite{Kim_SciRep2017, Kang_NanoLett2017}.

In Fig.~\ref{Fig_S_EDM_CB_VB}(b,c) photoemission peak maxima of the conduction band for both $\Sigma$ to ${\rm K}$ valleys shift towards higher binding energies, away from the Fermi level.
However withing experimental broadening of corresponding spectra, one could not with absolute certainty conclude, that conduction band minima is shifting from $\Sigma$ to ${\rm K}$ valley. 
Therefore, despite obvious spectral intensity shift from $\Sigma$ to ${\rm K}$ valley in Fig.~\ref{Fig_S_CB}, we could not presume the crossover to a monolayer conduction band dispersion in alkali-metal dosed MoSe$_2$.
Correspondingly, in Fig.~\ref{Fig_S_EDM_CB_VB}(e,f) photoemission peak maxima of the valence band for both $\Gamma$ to ${\rm K}$ valleys shift towards Fermi level with dosing.
Withing experimental broadening of corresponding spectra, one could not determine, that valence band maxima is shifting from $\Gamma$ to ${\rm K}$ valley. 
The lower intensity surface related features might be especially well masked by the bulk component at $\Gamma$ valley as compared to ${\rm K}$ valley as seen in Fig.~\ref{Fig_S_EDC_VB}.
Additional broadening of the spectra due to significant k$_z$ dispersion at the centre of the Brillouin zone make distinguishing bulk and surface components there even more challenging.
We can not exclude electron doping induced indirect to direct band gap transition in MoSe$_2$, however this might be beyond "surface doping" regime achievable with alkali metal dosing at the surface.

\begin{figure}[hbt]
\centering
\includegraphics[width=\linewidth]{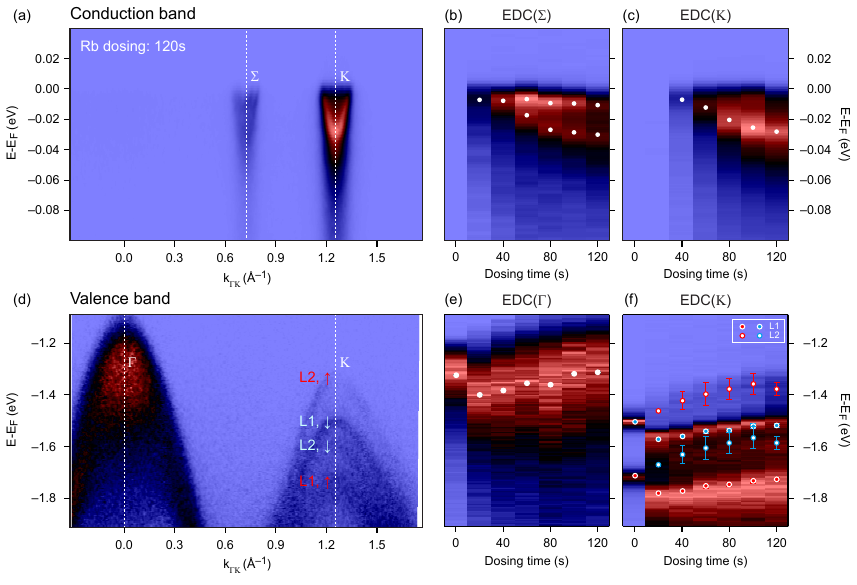}
\caption{\textbf{Evolution of the conduction band minima and valence band maxima with Rb dosing:}
(a) Conduction bands dispersion along $\Gamma$-${\rm K}$ direction;
(b) Dispersion of conduction band minima with surface doping obtained from EDC at $\Sigma$ point;
(c) Dispersion of conduction band minima with surface doping obtained from EDC at ${\rm K}$ point;
(d) Valence bands dispersion along $\Gamma$-${\rm K}$ direction. L1 and L2 notes topmost and second topmost layers, correspondingly;
(e) Dispersion of Valence band maxima with surface doping obtained from EDC at $\Sigma$ point;
(f) Dispersion of Valence band maxima with surface doping obtained from EDC at ${\rm K}$ point;}
\label{Fig_S_EDM_CB_VB}
\end{figure}

\begin{figure}[hbt]
\centering
\includegraphics[width=\linewidth]{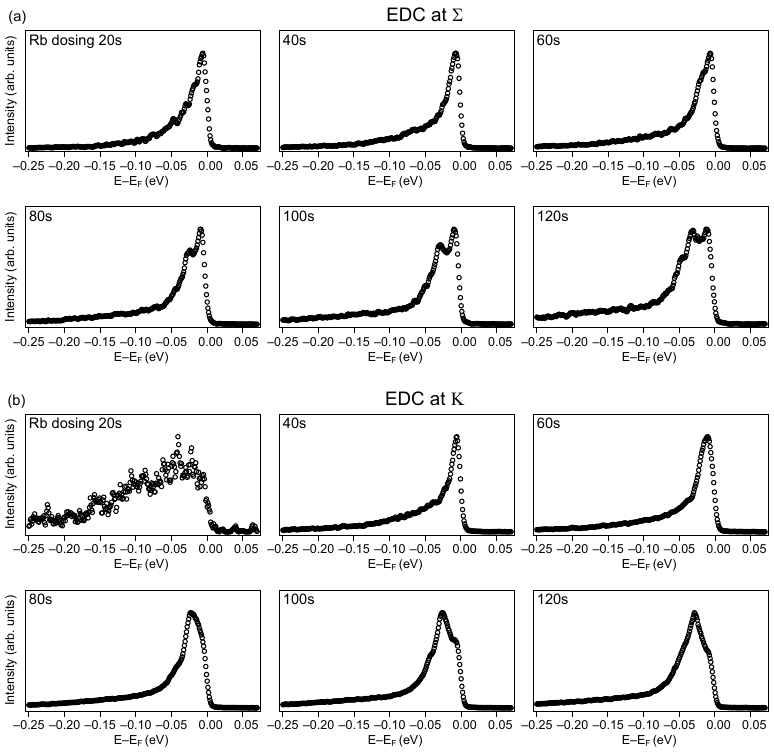}
\caption{\textbf{Evolution of conduction band minima with Rb dosing:}
(a) Energy distribution curves (EDCs) at $\Sigma$ point;
(b) Energy distribution curves (EDCs) at ${\rm K}$ point.}
\label{Fig_S_EDC_CB}
\end{figure}

\begin{figure}[hbt]
\centering
\includegraphics[width=\linewidth]{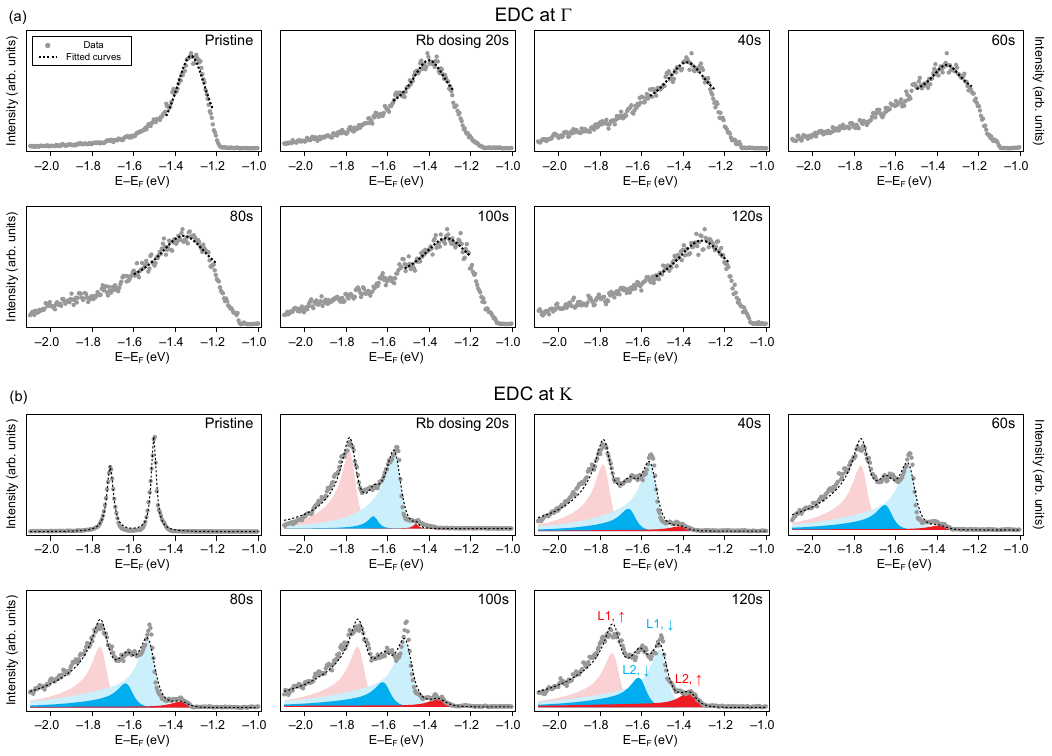}
\caption{\textbf{Evolution of valence band maxima with Rb dosing:}
(a) Energy distribution curves (EDCs) with fits at $\Gamma$ point;
(b) Energy distribution curves (EDCs) with fits at ${\rm K}$ point.}
\label{Fig_S_EDC_VB}
\end{figure}

\begin{figure}[hbt]
\centering
\includegraphics[width=\linewidth]{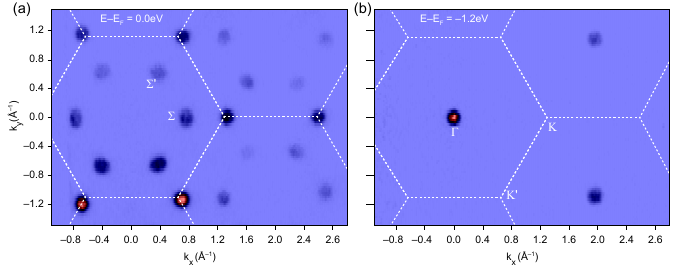}
\caption{\textbf{Constant energy maps (hv=180eV) of MoSe$_2$ surface after 140\,s Rb-dosing:}}
(a) Momentum distribution map at Fermi level and (b) Momentum distribution map at the top of the valence band.
\label{Fig_S_FSmaps}
\end{figure}

\clearpage
\subsection*{Rashba-like splitting at $\Sigma$ valley}
To analyse spin-splitting at $\Sigma$ valley as a function of surface doping we used Rashba-like model with $\Gamma$ as time-reversal invariant momenta (TRIM) point shown in Fig.~\ref{Fig_S_Rashba}(a).
\begin{figure}[h!]
\centering
\includegraphics[width=\linewidth]{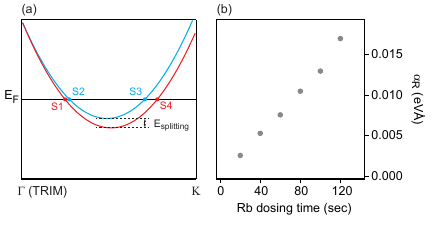}
\caption{\textbf{Spin-splitting at $\Sigma$ valley:
(a) Rashba-like model with $\Gamma$ as time-reversal invariant momentum (TRIM) point;
(b) Extracted Rashba parameter ($\alpha_R$).}}.
\label{Fig_S_Rashba}
\end{figure}
The effective Hamiltonian can be written as:
\begin{equation}
H(\vec{k})=\frac{\hbar^2}{2m^*}(\vec{k}-\vec{k_0})^2 \pm\alpha_R(\hat{n}\times\vec{\sigma})\cdot\vec{k}+E_{offset}
\end{equation}
where k$_0$ is the momentum of the conduction band minimum near $\Sigma$ without spin-orbit coupling, $\sigma$ is Pauli matrices, $E_{offset}$ is the chemical shift due to the electron doping via Rb dosing and m$^{*}$ is the effective mass.

From equation (1), one can make the system of equations with four $k_F$ (S1,S2,S3 and S4) shown in Fig.~\ref{Fig_S_Rashba}.
\begin{eqnarray}
\frac{\hbar^2}{2m^*}(k_{S1}-k_0)^2 -\alpha_{R}k_{S1}+E_{offset}=0 \\
\frac{\hbar^2}{2m^*}(k_{S2}-k_0)^2 +\alpha_{R}k_{S2}+E_{offset}=0 \\
\frac{\hbar^2}{2m^*}(k_{S3}-k_0)^2 +\alpha_{R}k_{S3}+E_{offset}=0 \\
\frac{\hbar^2}{2m^*}(k_{S4}-k_0)^2 -\alpha_{R}k_{S4}+E_{offset}=0
\end{eqnarray}
By subtracting the equation (5) with the equation (2), one can make
\begin{equation}
\frac{\hbar^2}{2m^*}(k_{S1}+k_{S4}-2k_0)-\alpha_R =0
\end{equation}
Likewise, one can subtract equation (4) from equation (3).
\begin{equation}
\frac{\hbar^2}{2m^*}(k_{S2}+k_{S3}-2k_0)+\alpha_R =0
\end{equation}
From equation (6) and (7), one can derive $k_0$ from four $k_F$ values
\begin{equation}
k_0 =\frac{k_{S1}+k_{S2}+k_{S3}+k_{S4}}{4}
\end{equation}
By replacing $k_0$ in equation (6) and (7), one can derive effective Rashba parameter $\alpha_{R}$
\begin{equation}
\alpha_R=\frac{\hbar^2}{2m^*}\frac{(k_{S1}+k_{S4}-k_{S2}-k_{S3})}{2}
\end{equation}
From equations (1), (8) and (9), the splitting energy $E_{splitting}$ can be derived as
\begin{equation}
E_{splitting}=2\alpha_Rk_0-\frac{2m^*\alpha_{R}^2}{\hbar^2}
\end{equation}

To fit the ARPES data, we used constant DFT value for the effective mass $m^{*}=0.54 m_e$ at $\Sigma$ valley and $k_F$ values obtained from MDC fits shown in Fig.~\ref{Fig_S_MDC_S}. From these $k_F$ values we estimate the experimental asymmetry of the Rashba splitting at $\Sigma$ valley of about $\sim$9\%.

\begin{figure}[h!]
\centering
\includegraphics[width=\linewidth]{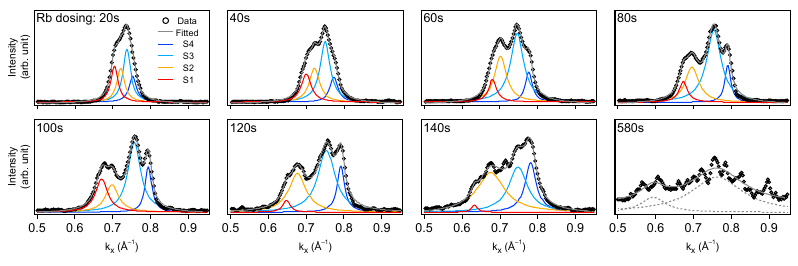}
\caption{\textbf{Momentum distribution curves (MDCs) of $\Sigma$ valleys at Fermi-level.}
}
\label{Fig_S_MDC_S}
\end{figure}

The effective Rashba parameter $\alpha_{R}$ versus Rb dosing time is shown in Fig.~\ref{Fig_S_Rashba}(b).
Corresponding extracted spin-split energy $E_{splitting}$ at $\Sigma$ valley as a function of Rb dosing time is shown in the main text Fig.~\ref{Fig_4}.

For comparison, for Au(111) and Cu(111) surface states measured Rashba parameter $\alpha_{R}$ = 330 and 38~meV{\AA} give energy splitting $E_{splitting}$ = 110 and 16~meV, correspondingly.

\clearpage
\end{widetext}
\end{document}